\documentclass[%
 preprint,
 amsmath,amssymb,
 aps,
]{revtex4-1}

\usepackage{graphicx}% Include figure files
\usepackage{dcolumn}% Align table columns on decimal point
\usepackage{bm}
\usepackage{mathtools}
\usepackage{breqn}
\usepackage{physics}
\usepackage{caption}
\usepackage{subcaption}
\usepackage{listings}
\usepackage[section]{placeins}
\usepackage{mathrsfs}

\begin{document}

%\preprint{APS/123-QED}

\title{	
Bianchi II and VII$_{h=0}$ Models Revisited Via The Euclidean-Signature Semi Classical Method
}% Force line breaks with \\

\author{Daniel Berkowitz}
 \altaffiliation{Physics Department, Yale University.\\ daniel.berkowitz@yale.edu \\ 217 Prospect St, New Haven, CT 06511
\\ This work is in memory of my parents, Susan Orchan Berkowitz, and Jonathan Mark Berkowitz\\ Mathematical Physics;Semi-Classical Methods;Quantum Cosmology}%Lines break automatically or can be forced with \\

\date{\today}% It is always \today, today,
             %  but any date may be explicitly specified

\begin{abstract}
We apply in a novel fashion a modified semi-classical method to the Bianchi II and VII$_{h=0}$ models when a cosmological constant, aligned electromagnetic field and stiff matter are present. Additionally we study the non-commutative quantum Bianchi II models when an aligned electromagnetic field is included. Through the use of the Euclidean-signature semi classical method we find a plethora of new solutions to these model's corresponding Lorentzian signature Wheeler DeWitt equations which we can interpret qualitatively. These new solutions for the aforementioned models involving matter sources reveal some potentially interesting effects that should be chronicled as possible phenomena that a toy model of quantum gravity can induce on the evolution of a quantum universe. Furthermore we find 'excited' states which behave differently from the 'excited' states of the Bianchi IX and Taub models that were previously uncovered using this method. By comparing and contrasting the 'excited' states given by these models we help facilitate a better understating of what constitutes an 'excited' state solution of the Wheeler Dewitt equation. Our results further show the utility of the Euclidean-signature semi classical method for tackling Lorentzian signature problems without having to invoke a Wick rotation. This feature of not needing to apply a Wick rotation makes this method potentially very useful for tackling a variety of problems in bosonic relativistic field theory and quantum gravity. 

\end{abstract}

\pacs{Valid PACS appear here}% PACS, the Physics and Astronomy
                             % Classification Scheme.
%\keywords{Suggested keywords}%Use showkeys class option if keyword
                              %display desired
\maketitle

%\tableofcontents

\section{\label{sec:level1}INTRODUCTION}
Since the early days\cite{misner1969quantum} of studying the symmetry reduced \cite{dewitt1967quantum} Wheeler DeWitt(WDW) equation, obtaining closed solutions for it of any kind has proven to be difficult. The "symmetry reduced" WDW equation is called such because it is derived by first reducing the infinite number of degrees of freedom present in general relativity to a finite number by utilizing the symmetries allowed in homogeneous space-times, such as Bianchi A models. When a diagonal Bianchi A metric of the following form
\begin{equation}\label{1}
d s^{2}=-N(t)^{2}d t^{2}+L^{2}e^{2 \alpha(t)}\left(e^{2 \beta(t)}\right)_{a b} \omega^{a} \omega^{b} 
\end{equation} is inserted into the Einstein-Hilbert action expressed in terms of the ADM\cite{arnowitt1959dynamical,arnowitt1962gravitation} formalism of general relativity 
\begin{equation}\label{2}
\mathcal{S}=\frac{c^3}{16\pi G}\int d t d^{3} x N \sqrt{h}\left(K_{a b} K^{a b}-K^{2}+R_3-2\Lambda\right) +\mathcal{S}_{matter};
\end{equation} the resulting ADM action has a finite number of degrees of freedom. 

In the above metric $L$ has units of length and together with $e^{3\alpha(t)}$ sets a scale for the spatial size of our cosmology. This can be seen because any shift in the scale factor $e^{\alpha(t)+\delta}$ where $\delta$ is a real number can be reabsorbed into $L$. Beyond inspecting (\ref{1}), it can be seen that $e^{\alpha(t)}$ acts as the scale factor for our models in these Misner variables by computing $\sqrt{-\operatorname{det} g}$ of the metric tensor (\ref{1}) expressed in orthonormal coordinates, yielding $e^{3\alpha(t)}$ which gives the relative spatial volume for a region of our Bianchi II and VII$_{h=0}$ universes.

From here one can construct a finite dimensional Hamiltonian which can be quantized as follows 
 \begin{equation}\label{3}
\begin{array}{l}{-e^{-3 \alpha} p_{\alpha}^{2} \longrightarrow \frac{\hbar^{2}}{e^{(3-B) \alpha}} \frac{\partial}{\partial \alpha}\left(e^{-B \alpha} \frac{\partial}{\partial \alpha}\right)} \\ {e^{-3 \alpha} p^{2}_{\beta_{+}}\longrightarrow \frac{-\hbar^{2}}{e^{3 \alpha}} \frac{\partial^{2}}{\partial \beta_{+}^{2}}} \\ {e^{-3 \alpha} p^{2}_{\beta_{-}} \longrightarrow \frac{-\hbar^{2}}{e^{3 \alpha}} \frac{\partial^{2}}{\partial \beta_{-}^{2}}}\end{array},
\end{equation} where B can be any real number and is the Hartle Hawking semi-general operator ordering parameter\cite{hartle1983wave}. 

In the notation introduced by Misner\cite{misner1969quantum,misner1969mixmaster} $e^{\alpha\left(t\right)}$ is a measure of the local scale factor of the spatial surface as we previously mentioned and 
\begin{equation}\label{4}
\left(e^{2 \beta(t)}\right)_{i j}=\operatorname{diag}\left(e^{2 \beta\left(t\right)_{+}+2 \sqrt{3} \beta\left(t\right)_{-}}, e^{2 \beta\left(t\right)_{+}-2 \sqrt{3} \beta\left(t\right)_{-}}, e^{-4 \beta\left(t\right)_{+}}\right),    
\end{equation}
where $\beta_+$ and $\beta_-$ measure the amount of anisotropy present on the spatial hypersurface. The $\omega^{i}$ factors are one forms defined on the spatial hypersurface of each Bianchi cosmology and obey $ d \omega^{i}=\frac{1}{2} C_{j k}^{i} \omega^{j} \wedge \omega^{k}$ where the $C_{j k}^{i}$ are the structure constants of the invariance Lie group associated with each particular class of Bianchi models.

The Bianchi II and VII$_{h=0}$ models which we will focus on in this paper have the following one forms respectively 
\begin{equation}\label{5}
\begin{aligned} 
\omega^{1} &=d y +x d z \\ \omega^{2} &=d z \\ \omega^{3} &= d x .
\end{aligned}
\end{equation}

\begin{equation}\label{6}
\begin{aligned} 
\omega^{1} &=\cos(z) d x+\sin(z) d y \\ \omega^{2} &=-\sin(z) d x+\cos(z) d y \\ \omega^{3} &= d z .
\end{aligned}
\end{equation}

Using the methodology presented in \cite{uggla1995classifying, waller1984bianchi,ryan1982bianchi} for writing out cosmological potentials for diagonal Bianchi A models with certain matter sources such as a cosmological constant and a primordial aligned electromagnetic field, we will use the Euclidean-signature semi classical method\cite{moncrief2014euclidean,marini2019euclidean} to study the following Lorentizian signature Wheeler DeWitt (WDW) equations\cite{aguero2007noncommutative}
\begin{equation}\label{7}
\begin{aligned} 
& \square \psi-B \frac{\partial \psi}{\partial \alpha}+U_{i} \psi=0 \\ &
U_{II}= \frac{1}{12} e^{4\alpha + 4\beta_+ +4\sqrt{3}\beta_-} +24\Lambda e^{6\alpha}+2b^{2}e^{2\alpha + 2\beta_+ +2\sqrt{3}\beta_-}+\rho \\& 
U_{VII}= \frac{4}{3} e^{4 (\alpha+\beta_+)} \sinh ^2\left(2 \sqrt{3} 
\beta_-\right) +24\Lambda e^{6\alpha}+2b^{2}e^{2\alpha + 2\beta_+ +2\sqrt{3}\beta_-}+\rho;
\end{aligned}
\end{equation}
where $\square$ is the three-dimensional d'Alembertian in the minisuperspace with signature, $(+--)$, $\rho$ is a constant which is our stiff matter term, and $b^{2}$ is a constant which represents how strong the aligned electromagnetic field is. 

 Even though on large scales our universe is incredibly isotropic and homogeneous it is highly likely that our early universe possessed a considerable degree of anisotropy and inhomogeneity which originated from quantum fluctuations within particle fields whose size was comparable to the primordial cosmological horizon. Thus it is useful to study anistropic or inhomogeneous classical/quantum cosmologies so we can better understand what our universe could have been like when it was extremely young. As a result the WDW equation is an excellent candidate to apply the Euclidean-signature semi classical method to in order to evaluate its efficacy. It already has been successfully applied to the Taub, Bianchi IX and VIII models\cite{berkowitz2020towards,berkowitz2021bianchi,bae2015mixmaster} and now we will expand that list to include the Bianchi II and  VII$_{h=0}$ models with matter sources
 
The wave functions we will obtain are easiest to interpret in a qualitative manner when we allow $\Lambda < 0$. Choosing a negative cosmological constant does not necessarily make our solutions non-physical. Recently there has been some interesting work\cite{biswas2009inflation,hartle2012accelerated,mithani2013inflation, hartle2014quantum,visinelli2019revisiting} done in studying inflation with a negative cosmological constant and the connection between  asymptotically Euclidean AdS wave functions to classical cosmological histories which exhibit phenomenology that one expects from universes with a positive cosmological constant. Furthermore, because our method is able to naturally relate solutions of Euclidean-signature equations, which can be used to construct Euclidean quantum cosmological models, to Lorentzian signature equation without having to invoke a Wick rotation or some notion of analytical continuation, it can provide a way to realize quantum domain-wall/cosmology correspondence\cite{mcfadden2010holography}. Thus from a theoretical point of view it is worthwhile to study quantum cosmologies that possess a negative cosmological constant. 

In addition we will only consider an electromagnetic field in which the electric and magnetic components of it are parallel with each other and are both non-vanishing. The case when only a primordial magnetic field is present can be realized in the quantum Bianchi I models \cite{ryan1982bianchi} whose 'ground' and 'excited' states can in theory be thoroughly analyzed with the method we will employ in this paper. Despite these restrictions the wave functions we obtain exhibit fascinating properties. As a result they provide further incentives to study wave functions of the universe derived from more general electric/magnetic field configurations and expand upon the mathematical results we will present in this paper. 

Beyond theoretical considerations, we include an aligned electromagnetic field in our WDW equation because new evidence\cite{neronov2010evidence,tavecchio2010intergalactic} for the existence of a femto Gauss strength intergalactic magnetic field has been uncovered by observing gamma rays. This provides further reasons to continue \cite{kamenshchik1993fermions,esposito1995relativistic,kobayashi2019early,jimenez2009cosmological,louko1988quantum,karagiorgos2018quantum,pavvsivc2012wheeler} studying electric/magnetic fields through the lens of quantum cosmology. Through studying the effects of electric/magnetic fields on quantum universes we can potentially better understand how seeds of anisotropy developed in our early universe which we can observe\cite{bennett2013nine,hinshaw2013nine,ade2016planck} today in the CMB. Recent work\cite{kahniashvili2001cmb,paoletti2011cmb,miyamoto2014cmb,hortua2017reduced} has been conducted in trying to determine what signatures a primordial magnetic field would induce on the various spectrums which can be derived from the CMB. By studying what effects aligned electromagnetic fields can induce on  our commutative/non-commutative Bianchi II and commutative VII$_{h=0}$ wave functions we can contribute to the theoretical portion of that task.

Before we proceed we should discuss the problem of time. Equations (\ref{7}) can be seen as the analogue of the Schr$\text{\" o}$dinger equation for these quantum cosmologies. They however possess many fundamental differences from the Schr$\text{\" o}$dinger equation which obscures the meaning behind $\psi$. Two notable differences are the absence of any first order time derivative, and the requirement that physically meaningful $\psi$'s must be annihilated by the quantized Hamiltonian constraint $\hat{\mathcal{H}}$, which leads to the problem of time manifesting itself as
\begin{equation}\label{8}
\begin{aligned}
& i \hbar \frac{\partial \Psi}{\partial t}=N \hat{\mathcal{H}}_{\perp} \Psi \\ &
\frac{\partial \Psi}{\partial t}=0.
\end{aligned}
\end{equation}
A way around this for our purposes is to denote one of the Misner variables to be our clock. A good clock increases monotonically. Out of the variables we can choose from, $\alpha$ which is related to the spatial size of our Bianchi II universe is the best candidate for our clock and will be for practical purposes our "time" \cite{dewitt1967quantum} parameter.

This paper will be organized as follows. In the next section we will explain what the Euclidean-signature semi classical method is and how it applies to quantum diagonal Bianchi A models and other physical theories. Then we will showcase our method by obtaining closed form solutions to the Bianchi II WDW equation that are similar to the ones first derived by \cite{obregon1996psi}. Afterwards we will show how all of the equations that this method provides can be solved which will give us a family of asymptotic and closed form solutions. From there we will derive the electromagnetic potential term which appears in (\ref{7}). Then we will obtain new solutions to the Bianchi II WDW equation when a cosmological constant, primordial aligned electromagnetic field and stiff matter are present and discuss its 'excited' states. Next we will take a detour from applying our modified semi-classical method and turn our attention to the non-commutative quantum Bianchi II models with a primordial aligned electromagnetic field, and stiff matter. 

Moving on from Bianchi II we will turn our attention to the quantum Bianchi VII$_{h=0}$ models. Using the Euclidean-signature semi classical method we will first study its vacuum 'ground' and 'excited' states. Afterwards we will study its 'ground' states when matter sources are present. Once we have computed all of our wave functions we will interpret\cite{moncrief1991amplitude} them by their aesthetic characteristics. For example we will assume, as was done in \cite{garcia2002noncommutative}, that each visible peak which is present for our wave functions represents a geometric state a quantum universe can tunnel in and out of. Finally we will provide some concluding remarks.

\section{\label{sec:level1}The Euclidean-signature semi classical method} 

Closed form solutions \cite{obregon1996psi,aguero2007noncommutative,socorro2009scalar, karagiorgos2019quantum,moncrief1991amplitude} to the symmetry reduced WDW equations, including those of the Bianchi II and VII$_{h=0}$ models have been found using a plethora of methods, such as traditional semi-classical methods and via elementary separation of variables after performing a suitable coordinate change. However recently a modified semi-classical method was introduced which has proven to be potent tool for solving problems in quantum mechanics and quantum cosmology, and has some interesting features which make it an attractive tool to tackle problems in relativistic bosonic field theory and quantum gravity. 

The origins of the Euclidean-signature semi classical method can be traced to modifying microlocal methods\cite{cohen1999effective} which have historically been employed to study the Schr$\text{\" o}$dinger equation. The goal of these modifications was to develop a  semi-classical method that can be applied to problems in bosonic and non-abelian gauge field theories such as Yang Mills theory, whose phase space is infinite dimensional. With this development in mind a modified semi classical method was first applied\cite{moncrief2012modified} to a sequence of one dimensional nonlinear quantum harmonic oscillators. It was found that the eigenvalues estimated from this method agreed with those computed using standard Rayleigh/Schr$\text{\" o}$dinger perturbation theory\cite{zz}. However the wave functions computed via this modified semi-classical method more accurately captured the more-rapid-than-Gaussian decay known to hold for the exact solutions to these problems. Most importantly though it was shown that the techniques involved in applying this method could be generalized to certain field theories which possess an infinite dimensional phase space. 

This same method, after being slightly modified became the Euclidean-signature semi classical method and was applied by Joseph Bae to the quantum diagonal Bianchi IX WDW equation \cite{bae2015mixmaster}. Using it he was able to prove the existence of a smooth and globally defined asymptotic 'ground' state solution for any arbitrary Hartle-Hawking ordering parameter\cite{hartle1983wave}. In addition he was able to investigate leading order 'excited' states of the quantum Bianchi IX models. The ability of this modified semi-classical method to facilitate a definition of 'excited' states for finite dimensional constrained theories makes it a valuable tool in quantum cosmology. 

Our application of this method will be somewhat different from what was previously carried out by J.Bae. Because he was only considering a vacuum model there was no preferred length scale. As a result, in his mathematical analysis he was able to effectively reduce the degrees of freedom of the Bianchi IX minisuperspace model by one. Furthermore the solution\cite{moncrief1991amplitude} of the Euclidean-signature Hamilton Jacobi equation he was working with induced a specific flow in $\beta$ space. Notably the flow induced by the 'wormhole' solution of the Bianchi IX Euclidean-signature Hamilton Jacobi equation drives\cite{bae2015wormhole,bae2015mixmaster} the anisotropic $\beta_{+}(t)$ and $\beta_{-}(t)$ variables to the origin as $t\rightarrow \infty$, irrespective of their initial location in $\beta$ space, which we hold to be purely real. As a result, the equations corresponding to the quantum corrections for the Bianchi IX models were solved by integrating them along a flow induced in a subset of minisuperspace. This type of integration only converges if the variables being integrated vanish as $t\rightarrow \infty$. 

Our approach will differ from this because we are including matter sources which pick out preferred length scales. Also our solutions to the Euclidean-signature Hamilton Jacobi equation don't induce a flow in minisuperspace which drives the $\beta$ variables to the origins. Instead we will show that either the transport equations which correspond to our quantum corrections are directly solvable by picking a clever ansatz which reduces the task of solving them to an algebraic problem or that the sequence of transport equations terminates after a finite number of them have been solved. Thus our application of the Euclidean-signature semi classical can be considered a novel one. 

The Euclidean-signature semi classical method also has exciting applications for bosonic relativistic field theory and may be able to shed some light on the Yang Mills 'mass gap' problem. This is so because  this method possess an advantage over traditional perturbative techniques used to analyze field theories. Notably this modified semi-classical method doesn't require splitting the theory up into one portion which is linear(non-interacting) and another portion which is a nonlinear(interacting) perturbation. This allows the fully interacting nature of the field theory to be present at every level of its analysis. 

Furthermore, Euclidean-signature equations may be more amenable to known mathematical theorems and results than their Lorentzian signature counterparts. Thus a technique which allows one to naturally relate the solutions of Euclidean-signature equations to Lorentzian signature equations is immensely valuable. Such a method may be able to avoid the common pitfalls\cite{gibbons1979positive,gibbons1993euclidean} associated with the Euclidean path integral approach to quantum gravity. As a result this method may help us prove the existence of formal solutions to the functional Euclidean-signature Einstein-Hamilton-Jacobi equation. This equation is of great importance because any valid operator ordering of the Lorentzian signature functional WDW equation in the semi-classical limit reduces to the Einstein-Hamilton-Jacobi equation. Because this method is able to naturally relate\cite{moncrief2014euclidean} the solutions of the Euclidean-signature Einstein-Hamilton-Jacobi equation to the Lorentzian signature functional WDW without having to invoke a Wick rotation it has the potential to further our understanding of quantum gravity. More information on how the Euclidean-signature semi classical method applies to quantum cosmology, quantum gravity, and a plethora of field theories, including the Yang Mills 'mass gap' problem is provided here\cite{moncrief2012modified,moncrief2014euclidean, marini2016euclidean,marini2019euclidean,marini2020euclidean}

Our outline of this method will follow closely \cite{moncrief2014euclidean}. The method described in this section and its resultant equations can in principle be used to find solutions (closed and asymptotic) to a wide class of quantum cosmological models such as all of the diagonal Bianchi A, Kantowski Sachs models, and the FLRW models.  
 
The first step we will take in solving the Wheeler DeWitt equation is to introduce the ansatz
 \begin{equation}\label{9}
\stackrel{(0)}{\Psi}_{\hbar}=e^{-S_{\hbar} / \hbar}
\end{equation}
where $S_{\hbar}$ is a function of $\left(\alpha,\beta_+,\beta_-\right)$. We will rescale $S_{\hbar}$ in the following way  
\begin{equation}\label{10}
\mathcal{S}_{\hbar} :=\frac{G}{c^{3} L^{2}} S_{\hbar}
\end{equation}
where $\mathcal{S}_{\hbar}$ is dimensionless and admits the following power series in terms of this dimensionless parameter
\begin{equation}\label{11}
X :=\frac{L_{\text { Planck }}^{2}}{L^{2}}=\frac{G \hbar}{c^{3} L^{2}}.
\end{equation}
The series is given by 
\begin{equation}\label{12}
\mathcal{S}_{\hbar}=\mathcal{S}_{(0)}+X \mathcal{S}_{(1)}+\frac{X^{2}}{2 !} \mathcal{S}_{(2)}+\cdots+\frac{X^{k}}{k !} \mathcal{S}_{(k)}+\cdots
\end{equation},
and as a result our initial ansatz now takes the following form 
\begin{equation}\label{13}
\stackrel{(0)}{\Psi}_{\hbar}=e^{-\frac{1}{X} \mathcal{S}_{(0)}-\mathcal{S}_{(1)}-\frac{X}{2 !} \mathcal{S}_{(2)}-\cdots}
\end{equation}.
Substituting this ansatz into the Wheeler-DeWitt equation and
requiring satisfaction, order-by-order in powers of X leads immediately to the sequence of equations

\begin{equation}\label{14}
\begin{aligned}
&{\left(\frac{\partial \mathcal{S}_{(0)}}{\partial \alpha}\right)^{2}-\left(\frac{\partial \mathcal{S}_{(0)}}{\partial \beta_{+}}\right)^{2}-\left(\frac{\partial \mathcal{S}_{(0)}}{\partial \beta_{-}}\right)^{2}}+U=0
\end{aligned}
\end{equation}
\begin{equation}\label{15}
\begin{aligned}
& 2\left[\frac{\partial \mathcal{S}_{(0)}}{\partial \alpha} \frac{\partial \mathcal{S}_{(1)}}{\partial \alpha}-\frac{\partial \mathcal{S}_{(0)}}{\partial \beta_{+}} \frac{\partial \mathcal{S}_{(1)}}{\partial \beta_{+}}-\frac{\partial \mathcal{S}_{(0)}}{\partial \beta_{-}} \frac{\partial \mathcal{S}_{(1)}}{\partial \beta_{-}}\right] \\ & +B \frac{\partial \mathcal{S}_{(0)}}{\partial \alpha}-\frac{\partial^{2} \mathcal{S}_{(0)}}{\partial \alpha^{2}}+\frac{\partial^{2} \mathcal{S}_{(0)}}{\partial \beta_{+}^{2}}+\frac{\partial^{2} \mathcal{S}_{(0)}}{\partial \beta_{-}^{2}}=0,
\end{aligned}
\end{equation},
\begin{equation}\label{16}
\begin{aligned}
& 2\left[\frac{\partial \mathcal{S}_{(0)}}{\partial \alpha} \frac{\partial \mathcal{S}_{(k)}}{\partial \alpha}-\frac{\partial \mathcal{S}_{(0)}}{\partial \beta_{+}} \frac{\partial \mathcal{S}_{(k)}}{\partial \beta_{+}}-\frac{\partial \mathcal{S}_{(0)}}{\partial \beta_{-}} \frac{\partial \mathcal{S}_{(k)}}{\partial \beta_{-}}\right] \\ & {+k\left[B \frac{\partial \mathcal{S}_{(k-1)}}{\partial \alpha}-\frac{\partial^{2} \mathcal{S}_{(k-1)}}{\partial \alpha^{2}}+\frac{\partial^{2} \mathcal{S}_{(k-1)}}{\partial \beta_{+}^{2}}+\frac{\partial^{2} \mathcal{S}_{(k-1)}}{\partial \beta_{-}^{2}}\right]} \\ & + \sum_{\ell=1}^{k-1} \frac{k !}{\ell !(k-\ell) !}\Biggr(\frac{\partial \mathcal{S}_{(\ell)}}{\partial \alpha} \frac{\partial \mathcal{S}_{(k-\ell)}}{\partial \alpha}-\frac{\partial \mathcal{S}_{(\ell)}}{\partial \beta_{+}} \frac{\partial \mathcal{S}_{(k-\ell)}}{\partial \beta_{+}}  - \frac{\partial \mathcal{S}_{(\ell)}}{\partial \beta_{-}} \frac{\partial \mathcal{S}_{(k-\ell)}}{\partial \beta_{-}}\Biggl) =0
\end{aligned}
\end{equation}

We will refer to $\mathcal{S}_{(0)}$ in our WDW wave functions as the leading order term, which can be used to construct a semi-classical approximate solution to the Lorentzian signature WDW equation, and call $\mathcal{S}_{(1)}$ the first order term. The $\mathcal{S}_{(1)}$ term can also be viewed as our first quantum correction, with the other $\mathcal{S}_{(k)}$ terms being the additional higher order quantum corrections, assuming that they are smooth and globally defined. This is reflected in the fact that the higher order transport equations depend on the operator ordering used in defining the Wheeler Dewitt equation, which is an artifact of quantization. Additionally in some cases one can find a solution to the $\mathcal{S}_{(1)}$ equation which allows the $\mathcal{S}_{(2)}$ equation to be satisfied by zero. Then one can write down the following as a solution to the WDW equation for either a particular value of the Hartle-Hawking ordering parameter, or for an arbitrary ordering parameter depending on the $\mathcal{S}_{(1)}$ which is found.   

\begin{equation}\label{17}
\stackrel{(0)}{\Psi}_{\hbar}=e^{-\frac{1}{X} \mathcal{S}_{(0)}-\mathcal{S}_{(1)}}
\end{equation}.

This can be easily shown. Let's take $\mathcal{S}_{(0)}$ and $\mathcal{S}_{(1)}$ as arbitrary known functions which allow the $\mathcal{S}_{(2)}$ transport equation to be satisfied by zero, then the $k=3$ transport equation can be expressed as 
\begin{equation}\label{18}
{2\left[\frac{\partial \mathcal{S}_{(0)}}{\partial \alpha} \frac{\partial \mathcal{S}_{(3)}}{\partial \alpha}-\frac{\partial \mathcal{S}_{(0)}}{\partial \beta_{+}} \frac{\partial \mathcal{S}_{(3)}}{\partial \beta_{+}}-\frac{\partial \mathcal{S}_{(0)}}{\partial \beta_{-}} \frac{\partial \mathcal{S}_{(3)}}{\partial \beta_{-}}\right]}=0
\end{equation}
which is clearly satisfied by $\mathcal{S}_{(3)}$=0. The $\mathcal{S}_{(4)}$ equation can be written in the same form as (\ref{18}) and one of its solution is 0 as well, thus resulting in the $\mathcal{S}_{(5)}$ equation possessing the same form as (\ref{18}). One can easily convince oneself that this pattern continues for all of the $k\geq 3$ $\mathcal{S}_{(k)}$ transport equations as long as the solution of the $\mathcal{S}_{(k-1)}$ transport equation is chosen to be 0. Thus if an $\mathcal{S}_{(1)}$ exists which allows one to set the solutions to all of the higher order transport equations to zero the infinite sequence of transport equations generated by our ansatz truncates to a finite sequence of equations which allows us to construct a closed form wave function satisfying the WDW equation. Not all solutions to the $\mathcal{S}_{(1)}$ transport equation will allow the $\mathcal{S}_{(2)}$ transport equation to be satisfied by zero; however in our case, we were able to find $\mathcal{S}_{(1)}$'s which cause the $\mathcal{S}_{(2)}$ transport equation to be satisfied by zero, thus allowing one to set all of the solutions to the higher order transport equations to zero as shown above. This will enable us to construct new 'ground' state closed form solutions to the Lorentzian signature Bianchi II Wheeler Dewitt equation for arbitrary ordering parameter. It should be noted that using an alternate form\cite{giampieri1991new} of operator ordering for the WDW equation one can construct solutions to it using just the $\mathcal{S}_{(0)}$ term.

There is still some work that needs to be done to rigorously define 'excited' states within the contexts of this method. However the definition which has done a satisfactory job of describing 'excited' states for the Bianchi IX models\cite{bae2015mixmaster} and for the Taub models\cite{berkowitz2020towards} starts with the ansatz given below  
\begin{equation}\label{19}
{\Psi}_{\hbar}={\phi}_{\hbar} e^{-S_{\hbar} / \hbar}
\end{equation}
where $$
S_{\hbar}=\frac{c^{3} L^{2}}{G} \mathcal{S}_{\hbar}=\frac{c^{3} L^{2}}{G}\left(\mathcal{S}_{(0)}+X \mathcal{S}_{(1)}+\frac{X^{2}}{2 !} \mathcal{S}_{(2)}+\cdots\right)
$$
is the same series expansion as before and ${\phi}_{\hbar}$ can be expressed as the following series 
\begin{equation}\label{20}
{\phi_{\hbar}=\phi_{(0)}+X \phi_{(1)}+\frac{X^{2}}{2 !} \phi_{(2)}+\cdots+\frac{X^{k(*)}}{k !} \phi_{(k)}+\cdots}
\end{equation}
with X being the same dimensionless quantity as before. 
Inserting (\ref{19}) with the expansions given by (\ref{12}) and (\ref{20}) into the Wheeler DeWitt equation (\ref{7}) and by matching equations in powers of X leads to the following sequence of equations. 
\begin{equation}\label{21}
-\frac{\partial \phi_{(0)}}{\partial \alpha} \frac{\partial \mathcal{S}_{(0)}}{\partial \alpha}+\frac{\partial \phi_{(0)}}{\partial \beta_{+}} \frac{\partial \mathcal{S}_{(0)}}{\partial \beta_{+}}+\frac{\partial \phi_{(0)}}{\partial \beta_{-}} \frac{\partial \mathcal{S}_{(0)}}{\partial \beta_{-}}=0,
\end{equation},
\begin{equation}\label{22}
\begin{aligned}
&{-\frac{\partial \phi_{(1)}}{\partial \alpha} \frac{\partial \mathcal{S}_{(0)}}{\partial \alpha}+\frac{\partial \phi_{(1)}}{\partial \beta_{+}} \frac{\partial \mathcal{S}_{(0)}}{\partial \beta_{+}}+\frac{\partial \phi_{(1)}}{\partial \beta_{-}} \frac{\partial \mathcal{S}_{(0)}}{\partial \beta_{-}}} \\ & {+\left(-\frac{\partial \phi_{(0)}}{\partial \alpha} \frac{\partial \mathcal{S}_{(1)}}{\partial \alpha}+\frac{\partial \phi_{(0)}}{\partial \beta_{+}} \frac{\partial \mathcal{S}_{(1)}}{\partial \beta_{+}}+\frac{\partial \phi_{(0)}}{\partial \beta_{-}} \frac{\partial \mathcal{S}_{(1)}}{\partial \beta_{-}}\right)} \\ & {+\frac{1}{2}\left(-B \frac{\partial \phi_{(0)}}{\partial \alpha}+\frac{\partial^{2} \phi_{(0)}}{\partial \alpha^{2}}-\frac{\partial^{2} \phi_{(0)}}{\partial \beta_{+}^{2}}-\frac{\partial^{2} \phi_{(0)}}{\partial \beta_{-}^{2}}\right)=0,}
\end{aligned}
\end{equation}
\begin{equation}\label{23}
\begin{aligned}
& -\frac{\partial \phi_{(k)}}{\partial \alpha} \frac{\partial \mathcal{S}_{(0)}}{\partial \alpha}+\frac{\partial \phi_{(k)}}{\partial \beta_{+}} \frac{\partial \mathcal{S}_{(0)}}{\partial \beta_{+}}+\frac{\partial \phi_{(k)}}{\partial \beta_{-}} \frac{\partial \mathcal{S}_{(0)}}{\partial \beta_{-}} \\ & 
+k\Biggr(-\frac{\partial \phi_{(k-1)}}{\partial \alpha} \frac{\partial \mathcal{S}_{(1)}}{\partial \alpha}+\frac{\partial \mathcal{S}_{(1)}}{\partial \beta_{+}} \frac{\partial \mathcal{S}_{(1)}}{\partial \beta_{+}}+\frac{\partial \phi_{(k-1)}^{(*)}}{\partial \beta_{-}} \frac{\partial \mathcal{S}_{(1)}}{\partial \beta_{-}}\Biggr) \\ & 
+\frac{k}{2}\Biggr(-B \frac{\partial \phi_{(k-1)}}{\partial \alpha}+\frac{\partial^{2} \phi_{(k-1)}^{(*)}}{\partial \alpha^{2}}-\frac{\partial^{2} \phi_{(k-1)}}{\partial \beta_{+}^{2}}-\frac{\partial^{2} \phi_{(k-1)}}{\partial \beta_{-}^{2}}\Biggr)\\ & -
\sum_{\ell=2}^{k} \frac{k !}{\ell !(k-\ell) !}\Biggr( \frac{\partial \phi_{(k-\ell)}}{\partial \alpha} \frac{\partial \mathcal{S}_{(\ell)}}{\partial \alpha}-\frac{\partial \phi_{(k-\ell)}}{\partial \beta_{+}} \frac{\partial \mathcal{S}_{(\ell)}}{\partial \beta_{+}} -  \frac{\partial \phi_{(k-\ell)}}{\partial \beta_{-}} \frac{\partial \mathcal{S}_{(\ell)}}{\partial \beta_{-}}\Biggr) =0.
\end{aligned}
\end{equation}

It can be seen from computing $\frac{d\phi_{(0)}\left(\alpha,\beta_+,\beta_-\right)}{dt}=\dot{\alpha}\frac{\partial \phi_{(0)}}{\partial \alpha}+\dot{\beta_+}\frac{\partial \phi_{(0)}}{\partial \beta_+}+\dot{\beta_-}\frac{\partial \phi_{(0)}}{\partial \beta_-}$, and inserting $\left(4.9, \hspace{1 mm} 4.18-4.20\right)$ from \cite{moncrief2014euclidean} that $\phi_{(0)}$ is a conserved quantity under the flow of $S_{0}$. This means that any function $F\left(\phi_{(0)}\right)$ is also a solution of equation $\left(21\right)$. Wave functions constructed from these functions of $\phi_{0}$ are only physical if they are smooth and globally defined. If we choose our $\phi_{0}$ to have the form $f\left(\alpha,\beta_+,\beta_-\right)^{m_{1}}g\left(\alpha,\beta_+,\beta_-\right)^{m_{2}}$ where $f\left(\alpha,\beta_+,\beta_-\right)$ and  $g\left(\alpha,\beta_+,\beta_-\right)$ are some functions which are conserved under the flow of $\mathcal{S}_{(0)}$ and vanish for some finite values of the Misner variables then we must restrict $m_{i}$ to be either zero or a positive integer. For bound states, $m_{i}$ can plausibly be interpreted as graviton excitation numbers \cite{bae2014quantizing}. This makes our 'excited' states akin to bound states in quantum mechanics like the harmonic oscillator whose excited states are denoted by discrete integers as opposed to a continuous index. This discretization of the quantities that denotes our 'excited' states is the mathematical manifestation of quantization one would expect excited states to possess in quantum dynamics. If our conserved quantities do not vanish in minisuperspace then our 'excited' states can be interpreted as 'scattering' states akin to the quantum free particle and $m_{i}$ can be any pair real number.

As we will show soon, these 'excited' state transport equations can be solved in an elegant manner for the vacuum Bianchi II models. Studying both the mathematical properties of our 'excited' state transport equations and of the vacuum Bianchi II models can shed light on how the Euclidean-signature semi classical method can be applied to other problems in physics such as bosonic field theory\cite{marini2020euclidean}. In a previous study\cite{bae2015quantizing} the perturbations of the LRS Bianchi IX models were quantized to obtain an interpretation of the Bianchi IX model's 'excited' states; it may prove useful to do the same for the LRS Bianchi II models to show that the 'excited' states that this method can provide are indeed actual excited states. Additional information about 'excited' states can be found in \cite{moncrief2014euclidean}. In what follows we will work in a set of units in which $X=1$, unless stated otherwise.

\section{\label{sec:level1}'Ground' States Of The Vacuum Bianchi II Wheeler DeWitt Equation } 

As was reported in \cite{obregon1996psi}, there are three solutions to the Euclidean-signature Hamilton Jacobi equation (\ref{14}) that corresponds to the Bianchi II WDW equation (\ref{7}) when $\Lambda=0$, $b=0$, and $\rho=0$ given by 
\begin{equation}\label{24}
\begin{aligned}
\mathcal{S}^{1}_{(0)}:= \frac{1}{12} e^{2 \left(\alpha+\sqrt{3} \beta_-+\beta_+\right)}.
\end{aligned}
\end{equation}

\begin{equation}\label{25}
\begin{aligned}
\mathcal{S}^{2}_{(0)}:=\left(\frac{1}{12} e^{2 \alpha+2 \sqrt{3} \beta_-+2 \beta_+}+f(2\alpha+2\beta_{+})\right)
\end{aligned}
\end{equation}

\begin{equation}\label{26}
\begin{aligned}
\mathcal{S}^{3}_{(0)}:=\left(\frac{1}{12} e^{2 \alpha+2 \sqrt{3} \beta_-+2 \beta_+}+ g(2 \alpha+\sqrt{3} \beta_--\beta_+)\right) 
\end{aligned}
\end{equation}
where f(x) and g(x) are arbitrary single variable functions. In this section we will study 'ground' state solutions to the Bianchi II Wheeler DeWitt equation obtained using the Euclidean-signature semi classical method. The superscripts for the $ \mathcal{S}_{(k)}$ terms in this paper will play the role of an index to keep track of them unless stated otherwise.

Starting with (\ref{24}) if we insert it into our first 'ground' state transport equation (\ref{15}) we obtain the following simple differential equation 

\begin{equation}\label{27}
\begin{aligned}
2 \frac{\partial \mathcal{S}_{(1)}}{\partial \alpha}-2 \sqrt{3} \frac{\partial \mathcal{S}_{(1)}}{\partial \beta_-}-2 \frac{\partial \mathcal{S}_{(1)}}{\partial \beta_+}+\text{B}+6=0.
\end{aligned}
\end{equation}
This transport equation in principle has infinitely many solutions. However for the purposes of trying to find a $ \mathcal{S}_{(1)}$ which will allow the solutions to the higher order  $k \geq 2$ transport equations to be satisfied by zero, we will choose the following to be our $ \mathcal{S}_{(1)}$

\begin{equation}\label{28}
\begin{aligned}
\mathcal{S}^{1}_{(1)}:=\frac{1}{2} \left(-\text{B}+2 \text{x2}+2 \sqrt{3} \text{x3}-6\right) \alpha + \text{x2}\beta_+ +\text{x3}\beta_-,
\end{aligned}
\end{equation}
where x2 and x3 are free parameters, which can be real or complex. 
We choose this linear form because when we insert it into the $ \mathcal{S}_{(2)}$ transport equation (\ref{16}), its source term can be made to vanish by adjusting our free parameters. To clearly illustrate this if we insert (\ref{28}) into the source term of the $ \mathcal{S}_{(2)}$ transport equation we obtain the following expression  

\begin{equation}\label{29}
\begin{aligned}
f(x2,x3) = \text{B}^2-4 \left(2 \sqrt{3} (\text{x2}-3) \text{x3}-6 \text{x2}+2 \text{x3}^2+9\right),
\end{aligned}
\end{equation}
which can be easily made to vanish by solving for one of its free parameters (x2,x3) such that f(x2,x3)=0. The constraint of f(x2,x3)=0 was first derived in \cite{obregon1996psi}. If we solve for x2, the following $ \mathcal{S}_{(1)}$ will allow all of the higher order transport equations to be satisfied by zero and will allow us to easily write out a closed form solution for any arbitrary ordering parameter B to the Bianchi II Wheeler DeWitt equation 

\begin{equation}\label{30}
\begin{aligned}
&\mathcal{S}^{1}_{(1)}:=\frac{1}{2} \left(-\text{B}+2 \text{x2}+2 \sqrt{3} \text{x3}-6\right) \alpha + \text{x2}\beta_{+}+\text{x3}\beta_{-} \\& x2=\frac{\text{B}^2-8 \text{x3}^2+24 \sqrt{3} \text{x3}-36}{8 \left(\sqrt{3}
   \text{x3}-3\right)}.
\end{aligned}
\end{equation}
This results in us obtaining one of the solutions compatible with the constraint that was first reported in \cite{obregon1996psi} to the Bianchi II WDW equation for any arbitrary ordering parameter B 

\begin{equation}\label{31}
\begin{aligned}
\psi=e^{\left(-\frac{x3 \left(2 \alpha+\sqrt{3} \beta_--\beta_+\right)}{\sqrt{3}}-\frac{1}{12} e^{2 \left(\alpha+\sqrt{3} \beta_-+\beta_+\right)}-\frac{\left(B^{2}+12\right)
   (\alpha+\beta_+)}{8 \left(\sqrt{3} x3-3\right)}+\frac{1}{2} \alpha (B+2)-2 \beta_+\right)}.
\end{aligned}
\end{equation}
A nice feature of this solution is that it still possesses the free parameter x3, which allows one to form a wide variety of wave functions from it using superposition. The above is a procedure that in some cases\cite{berkowitz2020towards} can be employed to find closed form 'ground' state solutions to the Wheeler DeWitt equation using the Euclidean-signature semi classical method. 

Moving on to (\ref{25}) if we choose $f$ to be $\text{x1}e^{2\alpha+2\beta_{+}}$ where x1 is a free parameter and apply the same procedure we obtain an $\mathcal{S}_{(1)}$ which possesses one free parameter

\begin{equation}\label{32}
\begin{aligned}
&\mathcal{S}^{2}_{(1)}:=\text{x2}\alpha +\sqrt{3} \beta_-+\left(\frac{\text{B}}{2}+\text{x2}\right)\beta_+,
\end{aligned}
\end{equation}
and when inserted into (\ref{16}) results in a source term $-\frac{1}{2}B^{2}-6$ which does not vanish for any real values of the ordering parameter, but does vanish when $B=2\sqrt{3}i$. However if we want a solution involving a real value of the ordering parameter we can use the Euclidean-signature semi classical method to construct an asymptotic solution to the Bianchi II Wheeler Dewitt equation for any arbitrary ordering parameter for this $\mathcal{S}_{(0)}$ (\ref{25}).

If we choose the following ansatz for our higher order $k \geq 2$ quantum corrections 

\begin{equation}\label{33}
\begin{aligned}
&\mathcal{S}^{2}_{(k)}:=g(\text{B})_{k} e^{\left(-2 \alpha (k-1)-2 (k-1) \left(\sqrt{3} \beta_-+\beta_+\right)\right)};
\end{aligned}
\end{equation}
and insert it into (\ref{16}) we can prove that the problem of solving the higher order transport partial differential equations reduces to solving a recurrence equation, where we are solving for some function $g(\text{B})_{k}$ of the Hartle-Hawking ordering parameter. The first step in our proof is to insert (\ref{33}) into the homogeneous portion of equation (\ref{16}) which results in the following expression
\begin{equation}\label{34}
\begin{aligned}
2(k-1) g(\text{B})_{k} e^{-2 (k-2) \left(\alpha+\sqrt{3} \beta_-+\beta_+\right)}.
\end{aligned}
\end{equation}
The next step is to rewrite the source terms of equation (\ref{16}) as follows

\begin{equation}\label{35}
\begin{aligned}
&For \hspace{1mm} k=2 \\ & 2\left[B \frac{\partial \mathcal{S}_{(1)}}{\partial \alpha}-\frac{\partial^{2} \mathcal{S}_{(1)}}{\partial \alpha^{2}}+\frac{\partial^{2} \mathcal{S}_{(1)}}{\partial \beta_{+}^{2}}+\frac{\partial^{2} \mathcal{S}_{(1)}}{\partial \beta_{-}^{2}}\right]+ 2\Biggr(\frac{\partial \mathcal{S}_{(1)}}{\partial \alpha} \frac{\partial \mathcal{S}_{(1)}}{\partial \alpha}-\frac{\partial \mathcal{S}_{(1)}}{\partial \beta_{+}} \frac{\partial \mathcal{S}_{(1)}}{\partial \beta_{+}}  - \frac{\partial \mathcal{S}_{(1)}}{\partial \beta_{-}} \frac{\partial \mathcal{S}_{(1)}}{\partial \beta_{-}}\Biggl)
\end{aligned}
\end{equation}

\begin{equation}\label{36}
\begin{aligned}
&For \hspace{1mm} k=3 \\ & 3\left[B \frac{\partial \mathcal{S}_{(2)}}{\partial \alpha}-\frac{\partial^{2} \mathcal{S}_{(2)}}{\partial \alpha^{2}}+\frac{\partial^{2} \mathcal{S}_{(2)}}{\partial \beta_{+}^{2}}+\frac{\partial^{2} \mathcal{S}_{(2)}}{\partial \beta_{-}^{2}}\right]+6\Biggr(\frac{\partial \mathcal{S}_{(1)}}{\partial \alpha} \frac{\partial \mathcal{S}_{(2)}}{\partial \alpha}-\frac{\partial \mathcal{S}_{(1)}}{\partial \beta_{+}} \frac{\partial \mathcal{S}_{(2)}}{\partial \beta_{+}}  - \frac{\partial \mathcal{S}_{(1)}}{\partial \beta_{-}} \frac{\partial \mathcal{S}_{(2)}}{\partial \beta_{-}}\Biggl)
\end{aligned}
\end{equation}

\begin{equation}\label{37}
\begin{aligned}
&For \hspace{1mm} k > 3 \\ & k\left[B \frac{\partial \mathcal{S}_{(k-1)}}{\partial \alpha}-\frac{\partial^{2} \mathcal{S}_{(k-1)}}{\partial \alpha^{2}}+\frac{\partial^{2} \mathcal{S}_{(k-1)}}{\partial \beta_{+}^{2}}+\frac{\partial^{2} \mathcal{S}_{(k-1)}}{\partial \beta_{-}^{2}}\right] \\ & +\sum_{\ell=2}^{k-2} \frac{k !}{\ell !(k-\ell) !}\Biggr(\frac{\partial \mathcal{S}_{(\ell)}}{\partial \alpha} \frac{\partial \mathcal{S}_{(k-\ell)}}{\partial \alpha}-\frac{\partial \mathcal{S}_{(\ell)}}{\partial \beta_{+}} \frac{\partial \mathcal{S}_{(k-\ell)}}{\partial \beta_{+}} - \frac{\partial \mathcal{S}_{(\ell)}}{\partial \beta_{-}} \frac{\partial \mathcal{S}_{(k-\ell)}}{\partial \beta_{-}}\Biggl) \\& +2k\Biggr(\frac{\partial \mathcal{S}_{(1)}}{\partial \alpha} \frac{\partial \mathcal{S}_{(k-1)}}{\partial \alpha}-\frac{\partial \mathcal{S}_{(1)}}{\partial \beta_{+}} \frac{\partial \mathcal{S}_{(k-1)}}{\partial \beta_{+}}  - \frac{\partial \mathcal{S}_{(1)}}{\partial \beta_{-}} \frac{\partial \mathcal{S}_{(k-1)}}{\partial \beta_{-}}\Biggl)
\end{aligned}
\end{equation}
As the reader can easily verify if we were to insert (\ref{33}) into the source terms (\ref{35}) and (\ref{36}), the resulting expressions would be some constants which are proportional to (\ref{33}), and thus would allow one to calculate the k=2 and k=3 quantum corrections by simply solving for $g(B)_{k}$ and inserting it back into (\ref{33}). To prove that this is the case for the higher order $k > 3$ quantum corrections all we need to do is insert our $\mathcal{S}^{2}_{(k)}$ and our linear $\mathcal{S}^{2}_{(1)}$ into (\ref{37}). Doing so yields the following amazing simplification

\begin{equation}\label{38}
\begin{aligned}
&-2 (k-2) k (\text{B}-6 k+12) g(\text{B})_{k-1} e^{-2 (k-2) \left(\alpha+\sqrt{3} \beta_-+\beta_+\right)} \\& + \sum_{\ell=2}^{k-2} \frac{k !}{\ell !(k-\ell) !}\Biggr(-12 (l-1) (k-l-1) g(\text{B})_l e^{-2 (k-2) \left(\alpha+\sqrt{3} \beta_-+\beta_+\right)} g(\text{B})_{k-l}\Biggl) \\& +2 (\text{B}+6) (k-2) k g(\text{B})_{k-1} e^{-2 (k-2) \left(\alpha+\sqrt{3} \beta_-+\beta_+\right)}.
\end{aligned}
\end{equation}
Putting this all together, and solving for $g(\text{B})_k$ results in

\begin{equation}\label{39}
\begin{aligned}
&  g(\text{B})_{k}=\frac{\sum _{l=2}^{k-2} -\frac{12 (l-1) k! (k-l-1) g(B)_l g(B)_{k-l}}{l!
   (k-l)!}}{2-2 k}-6 k(k-2)  g(B)_{k-1}.
\end{aligned}
\end{equation}

As the reader can see our infinite sequence of linear partial differential equations has become a recurrence relation for our higher order quantum corrections. A computer algebra system like Mathematica can easily compute the terms of this recurrence relation and as a result the $\mathcal{S}^{2}_{(k)}$ quantum corrections can in principle be obtained to any order k. The above calculation presents an alternative to \cite{bae2015mixmaster} for obtaining asymptotic solutions to the Wheeler DeWitt equation using the Euclidean-signature semi classical method.

We have constructed a method to obtain all of the $\mathcal{S}^{2}_{(k)}$ quantum corrections to the semi-classical wave function $\stackrel{(0)}{\Psi}_{\hbar}=e^{-\frac{1}{X} \mathcal{S}^{2}_{(0)}}$, and as a result are able to construct a wide variety of asymptotic solutions to the Bianchi II Wheeler DeWitt equation for any Hartle-Hawking ordering parameter

\begin{equation}\label{40}
\begin{aligned}
&\mathcal{S}^{2}_{(k, \hspace{1mm} k>3)}:= \left(\frac{\sum _{l=2}^{k-2} -\frac{12 (l-1) k! (k-l-1) g(B)_l g(B)_{k-l}}{l!
   (k-l)!}}{2-2 k}-6 k(k-2)  g(B)_{k-1}\right)e^{\left(-2 \alpha (k-1)-2 (k-1) \left(\sqrt{3} \beta_-+\beta_+\right)\right)} \\& \mathcal{S}^{2}_{(2)}:=\frac{1}{4} \left(\text{B}^2+12\right) e^{-2 \alpha-2 \left(\sqrt{3} \beta_-+\beta_+\right)}  \\& \mathcal{S}^{2}_{(3)}:=-\frac{9}{2} \left(\text{B}^2+12\right) e^{-4 \alpha-4 \left(\sqrt{3} \beta_-+\beta_+\right)}  \\& \stackrel{(0)}{\Psi}_{\hbar}=e^{-\frac{1}{X} \mathcal{S}^{2}_{(0 \hspace{1mm} \pm)}-\mathcal{S}^{2}_{(1)}-\frac{X}{2 !} \mathcal{S}^{2}_{(2)}-\frac{X^{2}}{3 !} \mathcal{S}^{2}_{(3)}-\sum_{k=4}^{\infty}\frac{X^{k-1}}{k !}\mathcal{S}^{2}_{(k)}}.
\end{aligned}
\end{equation}
It would be instructive  to see if through some manner of non trivial summation such as a Borel sum if the resulting asymptotic terms converge to some wave function which behaves in an interesting fashion. Regardless of the convergence properties of these terms, the fact that such an asymptotic solution can be found in the first place is remarkable. There are other problems in physics where such a technique for computing an asymptotic expansion can prove to be very useful. 

The explicit forms of the $\mathcal{S}^{2}_{(2)}$ and $\mathcal{S}^{2}_{(3)}$ quantum corrections shown above can be easily computed by the reader using (\ref{34}), (\ref{35}), and (\ref{36}). Our quantum corrections possess the important property that they decay as $\alpha$ grows. Because $\alpha$ is related to the spatial size of our Bianchi II universe, physically it makes sense that our quantum corrections become increasingly important the smaller our universe becomes, while conversely becoming negligible in the classical limit of $\alpha >>0$. Because our solutions are asymptotic we only need to sum up a finite number of terms to get a good approximation for the full wave function. As a result of our solutions being asymptotic we will qualitatively analyze the properties of the following wave functions which are composed from $\mathcal{S}^{2}_{(0 )}$, $\mathcal{S}^{2}_{(1)}$, and $\mathcal{S}^{2}_{(2)}$ 

\begin{equation}\label{41}
\begin{aligned}
\psi=e^{ \left(\frac{1}{24} \left(-3 \left(\text{B}^2+12\right) e^{-2 \left(\alpha+\sqrt{3}
   \beta_-+\beta_+\right)}-2 e^{2 (\alpha+\beta_+)} \left(e^{2 \sqrt{3} \beta_-}+12 i \text{x1}\right)-24 \text{x2}
   (\alpha+\beta_+)-24 \sqrt{3} \beta_--12 \beta_+ \text{B}\right)\right)},
\end{aligned}
\end{equation}
where we made x1 an imaginary number. Because both x1 and x2 are free parameters there are infinitely many different wave functions we can choose to analyze. To narrow things down for our purposes we will set x2 and the ordering parameter B equal to zero, and then form the following wave function $\Psi=\int^{\infty}_{-\infty} e^{-x1^2}\psi dx1$ based on the linearity of the WDW equation resulting in

\begin{equation}\label{42}
\begin{aligned}
\Psi=\sqrt{\pi } e^{\left(\frac{1}{12} \left(-18 e^{-2 \left(\alpha+\sqrt{3} \beta_-+\beta_+\right)}-e^{2 \left(\alpha+\sqrt{3} \beta_-+\beta_+\right)}-3 e^{4 (\alpha+\beta_+)}-12 \sqrt{3}
   \beta_-\right)\right)}.
\end{aligned}
\end{equation}

We will display three plots(figures 1, 2, and 3) of this wave function for different values of $\alpha$ below and discuss them qualitatively towards the end of this paper.

\begin{figure}
\centering
\begin{subfigure}{.4\textwidth}
  \centering
  \includegraphics[scale=.13]{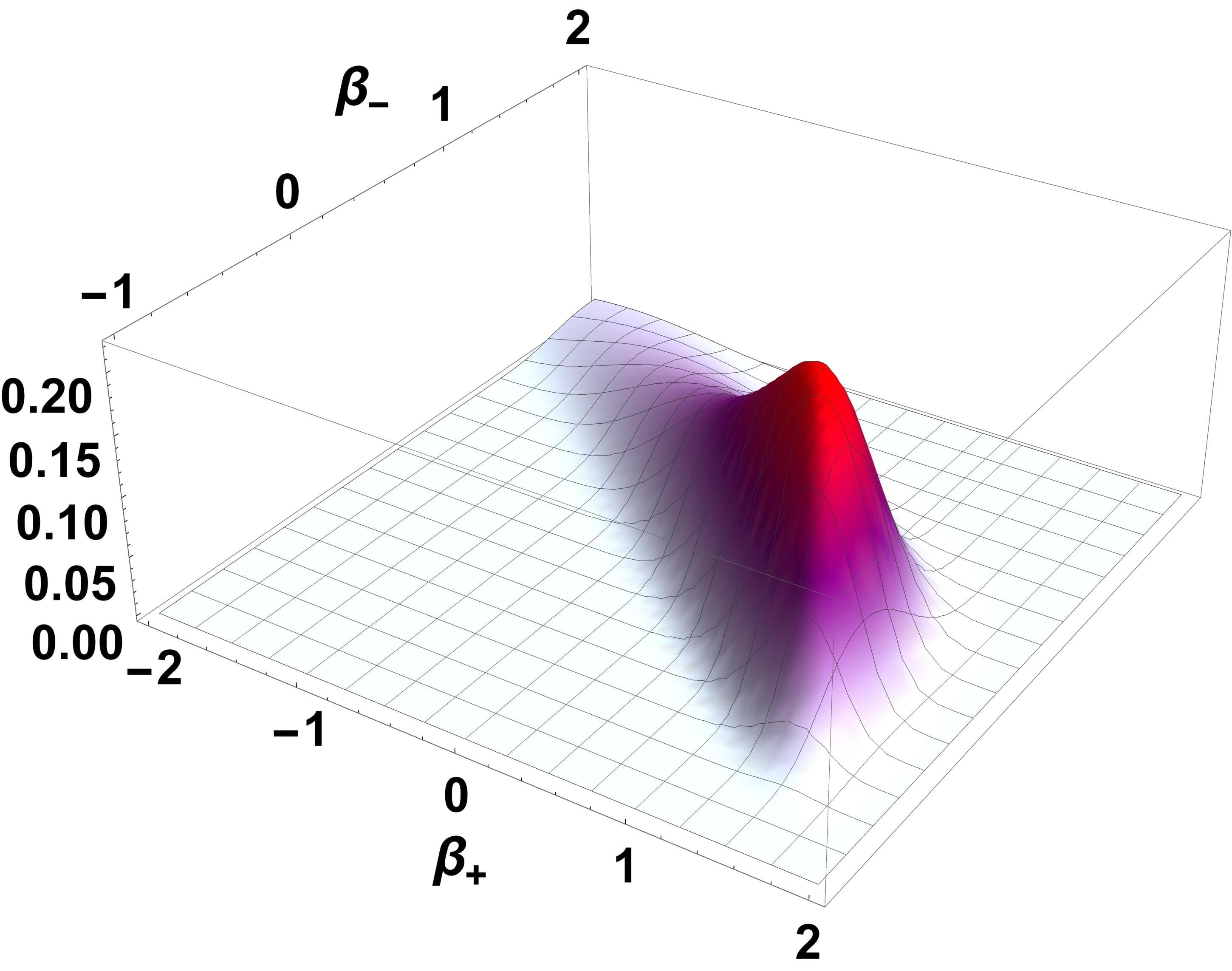}
  \caption{ $\alpha=-1.5$ \hspace{1mm} $x2=0$ \hspace{1mm} $B=0$}
  \label{1a}
\end{subfigure}%
\begin{subfigure}{.4\textwidth}
  \centering
  \includegraphics[scale=.13]{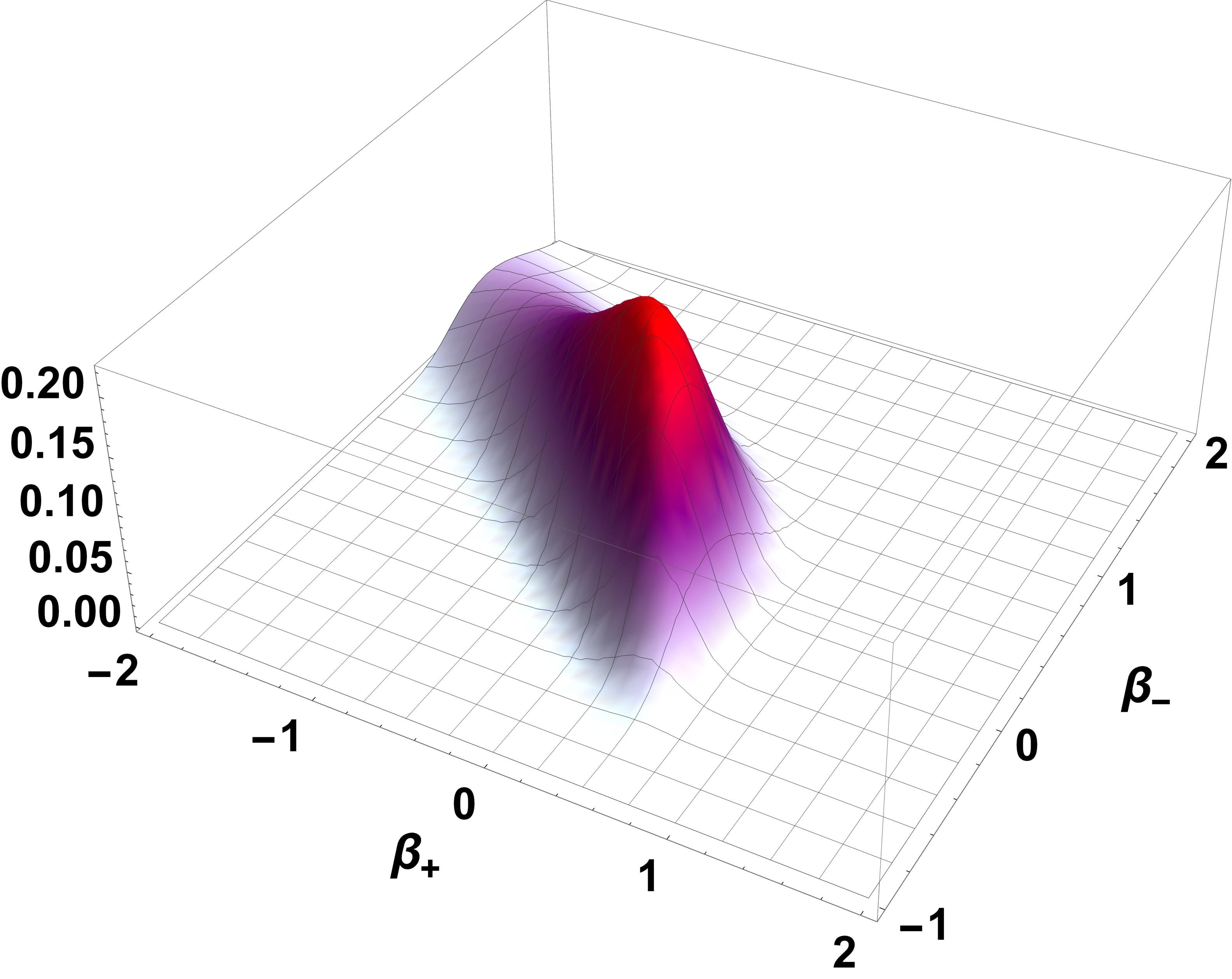}
  \caption{ $\alpha=0$ \hspace{1mm} $x2=0$ \hspace{1mm} $B=0$}
  \label{1b}
\end{subfigure}
\begin{subfigure}{.4\textwidth}
  \centering
  \includegraphics[scale=.13]{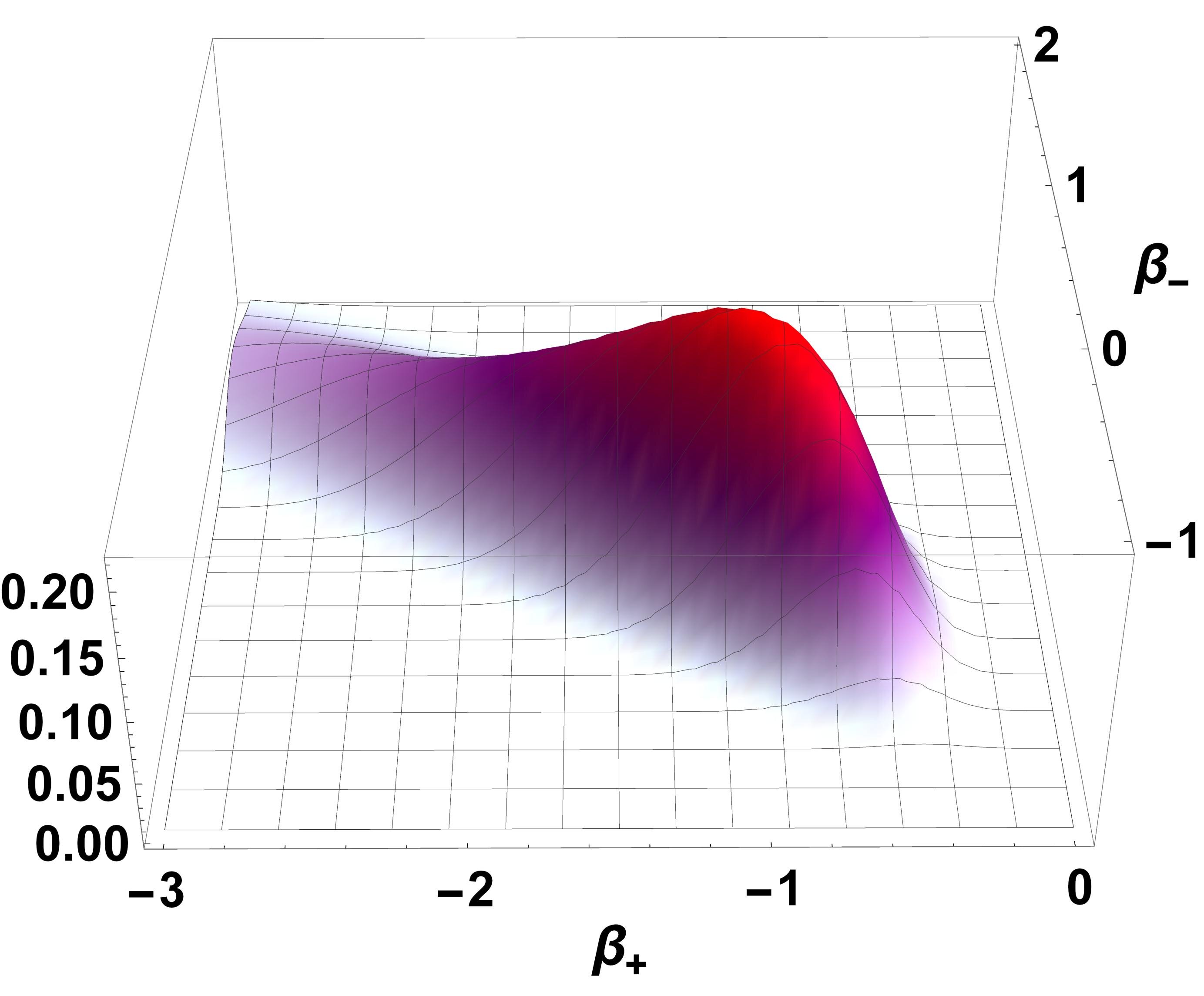}
  \caption{ $\alpha=1.5$ \hspace{1mm} $x2=0$ \hspace{1mm} $B=0$}
  \label{1c}
\end{subfigure}%
\caption{Plot of $\abs{\Psi}^{2}$ from (\ref{42}) for three different values $\alpha$.}
\label{fig 1}
\end{figure}

\section{\label{sec:level1}Solving The 'Excited' State Transport Equations In Closed Form } 

Even though the quantum Bianchi II models without any matter sources can be solved in closed form via separation of variables, applying this method has allowed us to obtain explicit quantum corrections to semi classical wave functions of the form (\ref{13}). For some cases such as the Taub models\cite{moncrief1991amplitude} superpositions of separable solutions can be constructed using integration which yield wave functions having the form of (\ref{13}). Computing an exact expression which looks like $\stackrel{(0)}{\Psi}_{\hbar}=e^{-\frac{1}{X} \mathcal{S}_{(0)}-\mathcal{S}_{(1)}-\frac{X}{2 !} \mathcal{S}_{(2)}-\cdots}$ though is contingent upon knowing how to integrate usually some Bessel function times a kernel such as $e^{-\omega^{2}}$ in closed form. As we have shown in the previous section our modified semi classical method allows us to bypass those mathematical difficulties and obtain solutions which are more mathematically transparent. In addition the wave functions that this method obtains for us possess non-trivial characteristics such as their behavior being highly dependent upon $\alpha$, which as was previously mentioned is our internal clock and also dictates the scale factor of our Bianchi II universes. Another non-trivial feature is the manifestation of discreteness in our wave functions which will be showcased in our 'excited states later on. 

We will now go over how all of the 'excited' state transport equations can be solved for the case when (\ref{24}) is our $\mathcal{S}_{(0)}$.
As it can be seen from our 'excited' state transport equations(\ref{21}-\ref{23}), in order to solve them we first need solutions to their ground state counterparts (\ref{14}-\ref{16}). If we insert (\ref{24}) into (\ref{21}) we obtain 
\begin{equation}\label{43}
\begin{aligned}
\left(\frac{\partial \phi_{(0)}}{\partial \alpha}-\sqrt{3} \frac{\partial \phi_{(0)}}{\partial \beta_-}-\frac{\partial \phi_{(0)}}{\partial \beta_+}\right)=0,
\end{aligned}
\end{equation}
which is an elementary linear transport equation which has the following solutions
\begin{equation}\label{44}
\begin{aligned}
\phi^{1}_{(0)}:= f_1\left( \left(3 \alpha+\sqrt{3} \beta_-\right),
   \left(3 \beta_+-\sqrt{3} \beta_-\right)\right),
\end{aligned}
\end{equation}
where $f_{1}$ is function of both the expressions $3 \alpha+\sqrt{3} \beta_-$ and $3 \beta_+-\sqrt{3} \beta_-$. As a result we have infinitely many choices for our $\phi_{(0)}$. We can exploit the properties of the 'excited' state transport equations and our solutions to the 'ground' state equations to pick an ansatz which will give us the forms for all of our $\phi_{(k)}$ terms. Using the same reasoning presented in \cite{moncrief2014euclidean} for the Bianchi IX models we will pick the following to be our ansatz for the higher order $\phi_{(k)}$ terms

\begin{equation}\label{45}
\begin{aligned}
\phi^{1}_{(k)}:=j(B)_{k}e^{ \left((m_{1}-2 k)\alpha +\frac{1}{\sqrt{3}} (-6 k+m_{1}-m_{2})\beta_-+ (m_{2}-2 k)\beta_+\right)}.
\end{aligned}
\end{equation}

The parameters $\left(m_1, m_2\right)$ in certain circumstances can plausibly be interpreted as graviton excitation numbers for the ultra long wavelength gravitational wave modes embodied in the $\left(\beta_+,\beta_-\right)$ anisotropic degrees of freedom \cite{bae2015quantizing}. If we assume $\left(m_1, m_2\right)$ represent physical quantities then they must be real numbers, despite the fact that states with complex $\left(m_1, m_2\right)$ can also satisfy the Wheeler DeWitt equation as will be shown below. Because our $\phi_{k}$ terms do not vanish anywhere, $m_{1}$ and $m_{2}$ can be any real numbers, and if they lead to excited states they would be scattering states. Before we solve for the explicit form of $j(B)_{k}$ we will pick a different ansatz to showcase the versatility of this method. 

If we choose our $\phi^{1}_{0}$ to be $(\alpha+\sqrt{3} \beta_-)^{m_{1}}$ $(3 \beta_+-\sqrt{3} \beta_-)^{m_{2}}$, we can obtain leading order bound states because both of our expressions vanish for real finite values of the Misner variables. Going beyond leading order we can actually find a closed form solution to the Bianchi II Wheeler Dewitt equation by simply inserting $(\alpha+\sqrt{3} \beta_-)^{m_{1}}$ $(3 \beta_+-\sqrt{3} \beta_-)^{m_{2}}e^{-\mathcal{S}^{1}_{(0)}-\mathcal{S}^{1}_{(1)}}$ into it and noticing that for m_{2}=1, m_{1}=0, and $x3=\frac{1}{4} \left(4 \sqrt{3}-\sqrt{\text{B}^2+12}\right)$ that it is satisfied by
\begin{equation}\label{46}
\begin{aligned}
\psi=\frac{1}{3} \left(3 \beta_+-\sqrt{3} \beta_-\right) e^{ \left(\frac{1}{12} \left(-e^{2 \left(\alpha+\sqrt{3} \beta_-+\beta_+\right)}+2 \alpha \left(2 \sqrt{3} \sqrt{\text{B}^2+12}+3
   \text{B}-6\right)+\sqrt{\text{B}^2+12} \left(3 \beta_-+\sqrt{3} \beta_+\right)-12 \left(\sqrt{3} \beta_-+\beta_+\right)\right)\right)}.
\end{aligned}
\end{equation}
This solution shows that one doesn't have to stick to the type of ans$\text{\" a}$tze used in \cite{moncrief2014euclidean} to find solutions to the Wheeler Dewitt equation using this method. As a matter of fact it may be more advantageous for the sake of finding 'excited' states for the vacuum Bianchi II models after choosing (\ref{24}) to be our $\mathcal{S}_{(0)}$ to use a different ansatz than (\ref{45}) as we will discuss soon. 

Moving on, if we pick (\ref{30}) to be the $\mathcal{S}_{(1)}$ for our 'excited' state transport equations, a significant simplification occurs. Because our closed form solution (\ref{31}) is constructed solely from an $\mathcal{S}_{(0)}$ and an $\mathcal{S}_{(1)}$ term all of the higher order $\mathcal{S}_{(k)}$ terms can be set to zero as was explained earlier. This significantly simplifies our 'excited' state transport equations because they depend on those higher order $\mathcal{S}_{(k)}$ terms which we can set to zero. The same is true for any Bianch A model which has a closed form solution where its $\mathcal{S}_{(k >1 )}$ terms vanish(an even greater simplification occurs if (\ref{7}) is satisfied by $e^{-\mathcal{S}_{(0)}}$).  As a result our sequence of transport equations becomes

\begin{equation}\label{47}
\begin{aligned}
& -\frac{\partial \phi_{(k)}}{\partial \alpha} \frac{\partial \mathcal{S}_{(0)}}{\partial \alpha}+\frac{\partial \phi_{(k)}}{\partial \beta_{+}} \frac{\partial \mathcal{S}_{(0)}}{\partial \beta_{+}}+\frac{\partial \phi_{(k)}}{\partial \beta_{-}} \frac{\partial \mathcal{S}_{(0)}}{\partial \beta_{-}} \\ & 
+k\Biggr(-\frac{\partial \phi_{(k-1)}}{\partial \alpha} \frac{\partial \mathcal{S}_{(1)}}{\partial \alpha}+\frac{\partial \mathcal{S}_{(1)}}{\partial \beta_{+}} \frac{\partial \mathcal{S}_{(1)}}{\partial \beta_{+}}+\frac{\partial \phi_{(k-1)}^{(*)}}{\partial \beta_{-}} \frac{\partial \mathcal{S}_{(1)}}{\partial \beta_{-}}\Biggr) \\ & 
+\frac{k}{2}\Biggr(-B \frac{\partial \phi_{(k-1)}}{\partial \alpha}+\frac{\partial^{2} \phi_{(k-1)}^{(*)}}{\partial \alpha^{2}}-\frac{\partial^{2} \phi_{(k-1)}}{\partial \beta_{+}^{2}}-\frac{\partial^{2} \phi_{(k-1)}}{\partial \beta_{-}^{2}}\Biggr).
\end{aligned}
\end{equation}
Our situation supremely simplifies further if we can find a $\phi_{k}$  which is able to satisfy its associated transport equation when it equals zero. If $\phi_{k}=0 $ satisfies the kth order 'excited' state transport equation then the k+1th order transport equation will reduce to 
\begin{equation}\label{48}
\begin{aligned}
-\frac{\partial \phi_{(k+1)}}{\partial \alpha} \frac{\partial \mathcal{S}_{(0)}}{\partial \alpha}+\frac{\partial \phi_{(k+1)}}{\partial \beta_{+}} \frac{\partial \mathcal{S}_{(0)}}{\partial \beta_{+}}+\frac{\partial \phi_{(k+1)}}{\partial \beta_{-}} \frac{\partial \mathcal{S}_{(0)}}{\partial \beta_{-}}=0, 
\end{aligned}
\end{equation}
which is satisfied by $\phi_{k+1}=0 $. Thus the k+2th order transport equations reduces to 
\begin{equation}\label{49}
\begin{aligned}
-\frac{\partial \phi_{(k+2)}}{\partial \alpha} \frac{\partial \mathcal{S}_{(0)}}{\partial \alpha}+\frac{\partial \phi_{(k+2)}}{\partial \beta_{+}} \frac{\partial \mathcal{S}_{(0)}}{\partial \beta_{+}}+\frac{\partial \phi_{(k+2)}}{\partial \beta_{-}} \frac{\partial \mathcal{S}_{(0)}}{\partial \beta_{-}}=0 
\end{aligned}
\end{equation}
and it is also satisfied by $\phi_{k+2}=0$. When a kth order $\phi_{k}$ equation is satisfied by zero, all of the higher order $\phi_{k+n}$ transport equations can also be satisfied by zero as well. This results in a truncation of the infinite sequence of 'excited' state transport equations to a finite sequence and allows one to find closed form solutions to the Wheeler DeWitt equation for any model to which the above applies to. Inserting our ansatz (\ref{45}) into (\ref{47}) yields

\begin{equation}\label{50}
\begin{aligned}
& j(\text{B})_{k-1} \Bigl(3 m_{1}\left(\text{B}^2+8 m_{2}+36\right)-3 m_{2}\left(\text{B}^2+16 m_{2}-36\right)+144 k^2
   \left(\sqrt{3} \text{x3}-3\right) \\& -144 k \left(\sqrt{3} \text{x3}-3\right)+24 m_{1}^2-8 \text{x3} (m_{1}+2 m_{2}) \left(\sqrt{3}
   m_{1}-\sqrt{3} m_{2}-3 \text{x3}+6 \sqrt{3}\right)\Bigr)\\& +24 \left(\sqrt{3} \text{x3}-3\right) j(\text{B})_k =0,
\end{aligned}
\end{equation}
which allows us to easily find a simple recurrence relation
 for $j(\text{B})_k$
\begin{equation}\label{51}
\begin{aligned}
& j(\text{B})_k=\frac{1}{{24 \left(\sqrt{3} \text{x3}-3\right)}}j(\text{B})_{k-1} \Bigl(3 m_{1}\left(\text{B}^2+8 m_{2}+36\right)-3
   m_{2}\left(\text{B}^2+16 m_{2}-36\right)+144 k^2 \left(\sqrt{3}
   \text{x3}-3\right) \\ &-144 k \left(\sqrt{3} \text{x3}-3\right)+24 m_{1}^2-8 \text{x3}
   (m_{1}+2 m_{2}) \left(\sqrt{3} m_{1}-\sqrt{3} m_{2}-3 \text{x3}+6
   \sqrt{3}\right)\Bigr).
\end{aligned}
\end{equation}

This recursion relation can be solved in closed form using Mathematica in terms of Pochhammer functions. The full expression is too long and cumbersome to express in this paper. However we will display the explicit form for $j(\text{B})_k$ and all of our $\phi_{k}$'s when our closed form solution (\ref{31}) has x3=0 
\begin{equation}\label{52}
\begin{aligned}
&j(\text{B})_k= \frac{1}{\pi}\Biggl((-6)^k \cos \left(\frac{1}{12} \pi  \sqrt{m_{1}\left(\text{B}^2+8 m_{2}+36\right)-m_{2}\left(\text{B}^2+16 m_{2}-36\right)+8
   m_{1}^2+36}\right) \\& \Gamma \left(k-\frac{1}{12} \sqrt{8 m_{1}^2+\left(\text{BB}^2+8 m_{2}+36\right) m_{1}-m_{2}
   \left(\text{B}^2+16 m_{2}-36\right)+36}+\frac{1}{2}\right) \\& \Gamma \left(k+\frac{1}{12} \left(\sqrt{8 m_{1}^2+\left(\text{B}^2+8
   m_{2}+36\right) m_{1}-m_{2}\left(\text{B}^2+16 m_{2}-36\right)+36}+6\right)\right)\Biggr) \\& \phi_{k}= j(\text{B})_ke^{ \left((m_{1}-2 k)\alpha +\frac{1}{\sqrt{3}} (-6 k+m_{1}-m_{2})\beta_-+ (m_{2}-2 k)\beta_+\right)}.
\end{aligned}
\end{equation}

Going back to (\ref{51}) and (\ref{45}) we see that we possess the freedom to pick the values of any of our free parameters m_{1}, m_{2}, and x3, for any value of k so that $\phi_{k}$=0. When $\phi_{k}$=0, the solutions to the subsequent transport equations can be satisfied by zero as well, thus truncating the infinite sequence of transport equations to a finite one; enabling us to construct closed form solutions using  $\mathcal{S}^{1}_{(0)}$,  $\mathcal{S}^{1}_{(1)}$, and $\phi_{0}$......$\phi_{k-1}$. Because for every value of k we can set our free parameters so that $\phi_{k}$=0, this enables us to construct a closed form solution using $\phi_{0}$......$\phi_{k-1}$, and because k can take on every possible positive integer value, we have found an infinite family of solutions to the Bianchi II Wheeler DeWitt equation for arbitrary ordering parameter. In addition because we only need to adjust one of our three free parameters so that $j(\text{B})_k$=0, each one of our solutions has two free parameters which we can vary. If we choose x3 to be the parameter which we adjust so that $j(\text{B})_k$=0, each one of our closed form solutions at a value of k can have their two 'excitation' numbers $m_{1}$ and $m_{2}$ be any real number. Naturally, this choice makes the most physical sense in terms of forming quantum Bianchi II scattering states. However, if we decide that $m_{1}$ and $m_{2}$ are not physical quantities, they can be complex numbers and still satisfy the Bianchi II Wheeler DeWitt equation.

Despite in principle solving all of the 'excited' state transport equations for a reasonable looking $\phi_{0}$, $\left(e^{3 \alpha+\sqrt{3} \beta_-}\right)^{m_{1}}\left(e^{3 \beta_+-\sqrt{3} \beta_-}\right)^{m_{1}}$, if one were to graph these wave functions it wouldn't be straightforward to interpret them as excited states unlike the Bianchi IX\cite{bae2015mixmaster} and Taub models\cite{berkowitz2020towards} which can be easily interpreted as excited states. Nonetheless the fact that we can solve these equations in the first place is a big mathematical feat and further shows the prowess of this method for proving the existence of solutions to Lorentzian signature problems. 

Furthermore this does not mean no 'excited' states exist for the vacuum Bianchi II models. In this section we chose to use (\ref{45}) as our $\phi_{k}$s because it allowed us to easily solve for all of the 'excited' state transport equations. Because our $\phi_{0}$ is a conserved quantity we could have used any function of it to construct our 'excited' states. There very well could exist a $\phi_{0}$ which results in wave functions that have (\ref{24}) as their semi classical term that qualitatively behave like 'excited' states. In addition the manifestation of 'excited' states for our choice of $\phi_{0}$ may be dependent on how we define our Hilbert space, which is task we delineate to a future work.  

If we choose a different $\mathcal{S}_{(0)}$ such as the infinitely many choices for (\ref{25}) and (\ref{26}) we may have obtained a simple form for our $\phi_{0}$ which could immediately lead to wave functions that behave as 'excited' states. Studying the perturbations of the LRS Bianchi II models as was done for the Taub models in \cite{bae2015quantizing} would also be very useful in establishing the existence of vacuum Bianchi II 'excited' states. As we will see though when matter sources are included, our solutions to the $\phi_{0}$ transport equation does result in wave functions which do behave as 'excited' states. 

\section{\label{sec:level1}Electromagnetic Potentials For Bianchi II and  VII$_{h=0}$}

In this section we will compare two methods for obtaining the WDW equation (\ref{7}). The first method will be based on directly quantizing the class of classical Hamilitonians for Bianchi A models that was developed in \cite{waller1984bianchi}. This will lead to a semi-classical treatment of our electromagnetic degree of freedom and will be what we use in the following sections to analyze how matter sources affect our wave functions. However we will also do a full quantum treatment of the electromagnetic degree of freedom and compare the two approaches. We will assume all of our electric and magnetic fields are parallel to each other  \cite{jacobs1970homogeneous}.

With this in mind our first task is to obtain solutions for Maxwell's equations in the space-time (\ref{1}) in terms of the Misner variables. In our calculations we will set $L= 1$ which has units of length. Starting from 
\begin{equation}\label{53}
\boldsymbol{A}=A_{0}dt+A_{1}\omega^{1}+A_{2}\omega^{3}+A_{3}\omega^{3}
\end{equation}
where $\omega^{0}=dt$,
and using the fact that $  d\omega^{i}=\frac{1}{2} C_{j k}^{i} \omega^{j} \wedge \omega^{k}$
to aide us in computing $\boldsymbol{F}=d\boldsymbol{A}= \frac{1}{2}F_{\mu v}\omega^{\mu} \wedge \omega^{v}$ results in the following expression for $F_{\mu v}$
\begin{equation}\label{54}
 F_{\mu v}=A_{v,\mu}-A_{ \mu,v}+A_{\alpha} C_{\mu v}^{\alpha}.
\end{equation}
In (\ref{54}) differentiation is done through a vector dual to our one forms $\omega^{\mu}$ which we denote as $X_{\mu}$. Thus $A_{v,\mu}=X_{\mu}A_{v}$. The electromagnetic portion of $\mathcal{S}_{matter}$ in (\ref{2})  is 
\begin{equation}\label{55}
\mathcal{S}_{matter}=\int dt dx^{3}N\sqrt{h}\left(-\frac{1}{16 \pi} F_{\mu \nu} F^{\mu \nu}\right),
\end{equation}
where 
\begin{equation}\label{56}
\begin{aligned}
h_{ab}=e^{2\alpha(t)}\operatorname{diag}\left(e^{2 \beta\left(t\right)_{+}+ 2\sqrt{3}\beta\left(t\right)_{-}}, e^{2 \beta\left(t\right)_{+}- 2\sqrt{3}\beta\left(t\right)_{-}}, e^{-4 \beta\left(t\right)_{+}}\right).
\end{aligned}
\end{equation}
Writing the action (\ref{55}) in terms of its vector potential $A$ and our structure constants results in the Lagrangian density which is derived in \cite{waller1984bianchi}
\begin{equation}\label{57}
\begin{aligned}
&\mathscr{L}=\Pi^{s} A_{0, s}-NH \\&
\mathscr{L}=\Pi^{s} A_{0, s}-\Pi^{s} A_{s, 0}-N\frac{2 \pi}{\sqrt{h}} \Pi^{s} \Pi^{p} h_{s p} \\& - \frac{N\sqrt{h}}{16 \pi} h^{i k} h^{s l}\left(2A_{[i,s]}+A_{m} C^{m}_{i s}\right)\left(2A_{[k,l]}+A_{m} C^{m}_{k l}\right),
\end{aligned}
\end{equation}
where
\begin{equation}\label{58}
\Pi^{s}=\frac{\partial \mathscr{L} }{\partial\left(X_{0} A_{s}\right)}=\frac{ h^{s j}\sqrt{h}}{4 N\pi} \left(-A_{0, j}+A_{j, 0}+A_{\alpha} C^{\alpha}_{0 j}\right),
\end{equation}
and we allow the shift $N^{k}$ to vanish. If we invoke the homogeneity of (\ref{1}) then we can say that $A_{i, j}=0$, and $ A_{0, j}=0$ which results in (\ref{57}) simplifying to 

\begin{equation}\label{59}
\mathscr{L}=\Pi^{s} A_{s, 0}-N\left[\frac{2 \pi}{\sqrt{h}} \Pi^{s} \Pi^{p} h_{p s}+\frac{\sqrt{h}}{16 \pi} h^{i k} h^{s l} C^{m}_{k l}C^{n}_{i s} A_{m} A_{n}\right].
\end{equation}
The non zero Bianchi II and  VII$_{h=0}$ structure constants are the following respectively

\begin{equation}\label{60}
\begin{aligned}
&C^{1}_{2 3}=-1 \\&
C^{1}_{3 2}=1 
\end{aligned}
\end{equation}

\begin{equation}\label{61}
\begin{aligned}
&C^{1}_{2 3}=-1 \\&
C^{1}_{3 2}=1 \\&
C^{2}_{1 3}=1 \\&
C^{2}_{3 1}=-1 
\end{aligned}
\end{equation}

We will now set $A_{2}$, $A_{3}$, $\Pi^{2}$, and $\Pi^{3}$ to zero as so our fields are aligned and only consider the electromagnetic field produced by $A_{1}$ and $\Pi^{1}$; doing so results in the following Lagrangian density
\begin{equation}\label{62}
\mathscr{L}=\Pi^{1} A_{1,0}-N\left[\frac{2 \pi}{\sqrt{h}} \Pi^{1} \Pi^{1} h_{1 1}+\frac{\sqrt{h}}{16 \pi} h^{i k} h^{s l} C^{1}_{k l}C^{1}_{i s} A_{1} A_{1}\right].
\end{equation}
As the reader can easily verify $ h^{i k} h^{s l} C^{1}_{k l}C^{1}_{i s}=\frac{2h_{11}}{h}$, which allows us to obtain the following set of Maxwell's equations when $A_{1}$ and $\Pi^{1}$ are varied
\begin{equation}\label{63}
\dot{A}_{1}-4 \pi \frac{1}{\sqrt{h}} \Pi^{1}h_{11}=0
\end{equation}
and 
\begin{equation}\label{64}
\dot{\Pi}^{1}+\frac{1}{4 \pi} \frac{1}{\sqrt{h}} h_{1 1} A_{1}=0.
\end{equation}
For the last equation we applied an integration by parts to the term $\Pi^{1} A_{1, 0}$ and dropped the total derivative term which vanishes at the spatial boundary. The solutions for (\ref{63}) and (\ref{64}) are 
\begin{equation}\label{65}
A_{1}=\sqrt{2}B_{0}cos(\theta(t))
\end{equation}
\begin{equation}\label{66}
\Pi^{1}=\frac{1}{2\sqrt{2}\pi}B_{0}sin(\theta(t))
\end{equation}
where $\theta(t)$ is an integral which is immaterial for our purposes and $B_{0}$ is an integration constant. Inserting (\ref{65}) and (\ref{66}) back into (\ref{62}) results in 
\begin{equation}\label{67}
\begin{aligned}
\mathscr{L}=\Pi^{1} A_{1,0}-N\frac{B_{0}^{2}}{4\pi\sqrt{h}}h_{11}\left(sin(\theta(t))^{2}+cos(\theta(t))^{2}\right)  =\Pi^{1} A_{1,0}-N\frac{B_{0}^{2}}{4\pi}e^{-\alpha(t)+2\beta(t)_{+}+2\sqrt{3}\beta(t)_{-}},
\end{aligned}
\end{equation}
From (\ref{59}) we can easily identify the electromagnetic Hamiltonian as 
\begin{equation}\label{68}
\begin{aligned}
H_{em}=\frac{B_{0}^{2}}{4\pi}e^{-\alpha(t)+2\beta(t)_{+}+2\sqrt{3}\beta(t)_{-}}
\end{aligned}
\end{equation}
which can be added to the Hamiltonian constraint which is derivable from our action (\ref{2}) and (\ref{3})
\begin{equation}\label{69}
\begin{aligned}
&e^{-3 \alpha(t)}\left(-p_{\alpha}^{2}+p_{+}^{2}+p_{-}^{2}\right)+U_{g}+\frac{B_{0}^{2}}{4\pi}e^{-\alpha(t)+2\beta(t)_{+}+2\sqrt{3}\beta(t)_{-}}+\rho=0
\end{aligned}
\end{equation}
Quantizing (\ref{69}) using the factor ordering we chose before, multiplying each side by $e^{3\alpha(t)}$, and rescaling $B_{0}$ results in the Wheeler DeWitt equations(\ref{7}). 

If we start with (\ref{62}) and directly quantize our component of the total Hamiltonian constraint which is proportional to the lapse N we obtain a similar, but slightly different contribution to the potential. Simplifying the term in brackets of (\ref{62}) results in the following contribution to the Hamiltonian constraint derived with (\ref{2})

\begin{equation}\label{70}
\begin{aligned}
H_{em}=N\left[\frac{e^{-\alpha+2\beta_{+}+2\sqrt{3}\beta_{-}}\left(16 \pi ^2 \Pi^{1}\Pi^{1}+A^{2}_{1}\right)}{8 \pi }\right].
\end{aligned}
\end{equation}
The term $\left(16 \pi ^2 \Pi^{1}\Pi^{1}+A^{2}_{1}\right)$ commutes with our total Hamiltonian constraint $H_{gravity} +H_{em}$. Thus we can solve the following WDW equation constructed by directly quantizing (\ref{70}) with the rest of our constraint 

\begin{equation}\label{71}
\square \Psi-B \frac{\partial \Psi}{\partial \alpha}-e^{2\alpha+2\beta_{+}+2\sqrt{3}\beta_{-}}\left(2\pi \frac{\partial^2 \Psi}{\partial A^{2}_{1}}+\frac{1}{8\pi}A^{2}_{1}\Psi\right)+U_{II \lor VII} \Psi=0
\end{equation}
by first solving this eigenvalue problem 
\begin{equation}\label{72}
-2\pi \frac{\partial^2 \Psi}{\partial A^{2}_{1}}+\frac{1}{8\pi}A^{2}_{1}\Psi=b_{n}\Psi.
\end{equation}
This is simply the Schr$\text{\" o}$dinger equation for a harmonic oscillator whose solutions are well known
\begin{equation}\label{73}
\begin{aligned}
&\Psi=\psi\left(\alpha,\beta_{+},\beta_{-}\right)e^{-\frac{A^{2}_{1}}{8 \pi }} H_{b_{n}-\frac{1}{2}}\left(\frac{A_{1}}{2 \sqrt{\pi }}\right)\\& b_{n}=\frac{1}{2}\left(1+2n\right).
\end{aligned}
\end{equation}
Inserting our $\Psi$ from (\ref{73}) into (\ref{71}) yields 
\begin{equation}\label{74}
\square \psi-B \frac{\partial \psi}{\partial \alpha}-b_{n}e^{2\alpha+2\beta_{+}+2\sqrt{3}\beta_{-}}+U_{II \lor VII} \psi=0.
\end{equation}
    This WDW equation is similar to what we had before except for the fact that the strength $b_{n}$ of the electromagnetic field is now quantized thanks to (\ref{73}). By first solving the classical $A_{i}$ equations (\ref{63}) in terms of the Misner variables we eliminate the electromagnetic field degree of freedom. By keeping it we can in theory study much more general quantum cosmologies for anisotropic models that involve electromagnetic fields. For now though working with just (\ref{69}) is sufficient for what will follow.

\section{\label{sec:level1}Closed Form Bianchi II 'Ground' States With Matter Sources } 

\begin{figure}
\centering
\begin{subfigure}{.4\textwidth}
  \centering
 \includegraphics[scale=.14]{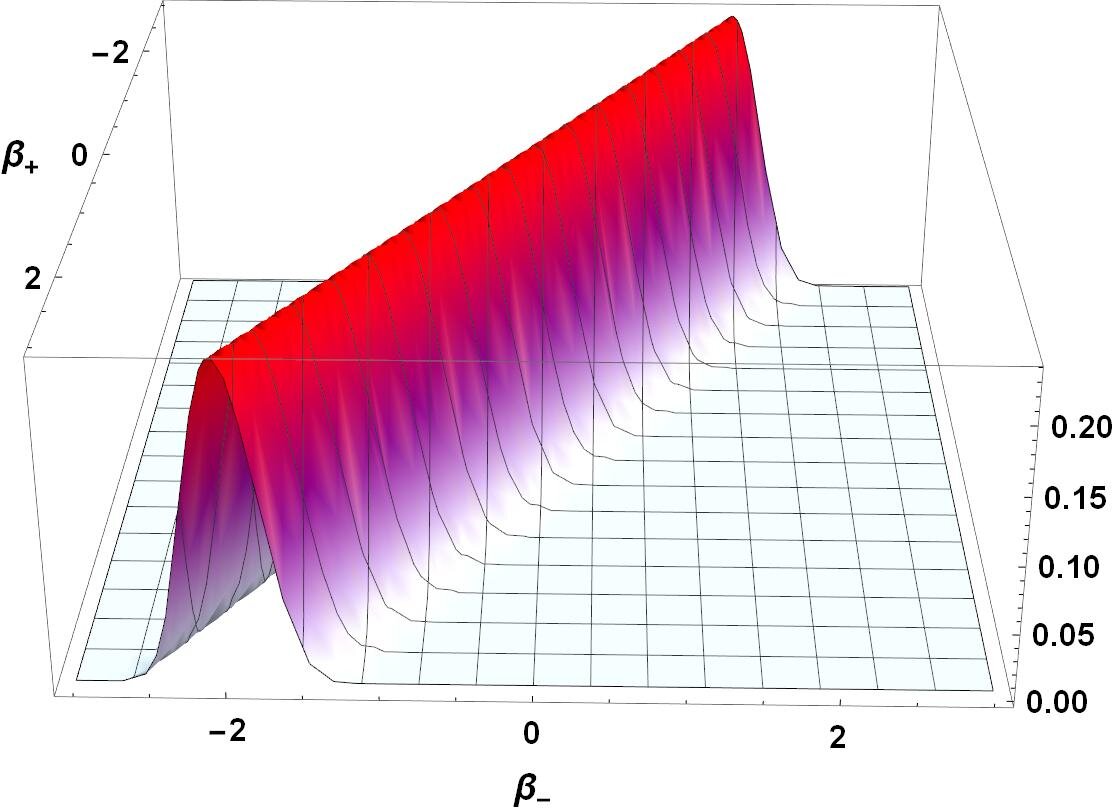}
\caption{$\alpha=-\frac{1}{4}$ $\hspace{1mm}$ $\Lambda=-1$ $\hspace{1mm}$ b=0 $\hspace{1mm}$ $\rho=0$}
  \label{2a}
\end{subfigure}%
\begin{subfigure}{.4\textwidth}
  \centering
 \includegraphics[scale=.14]{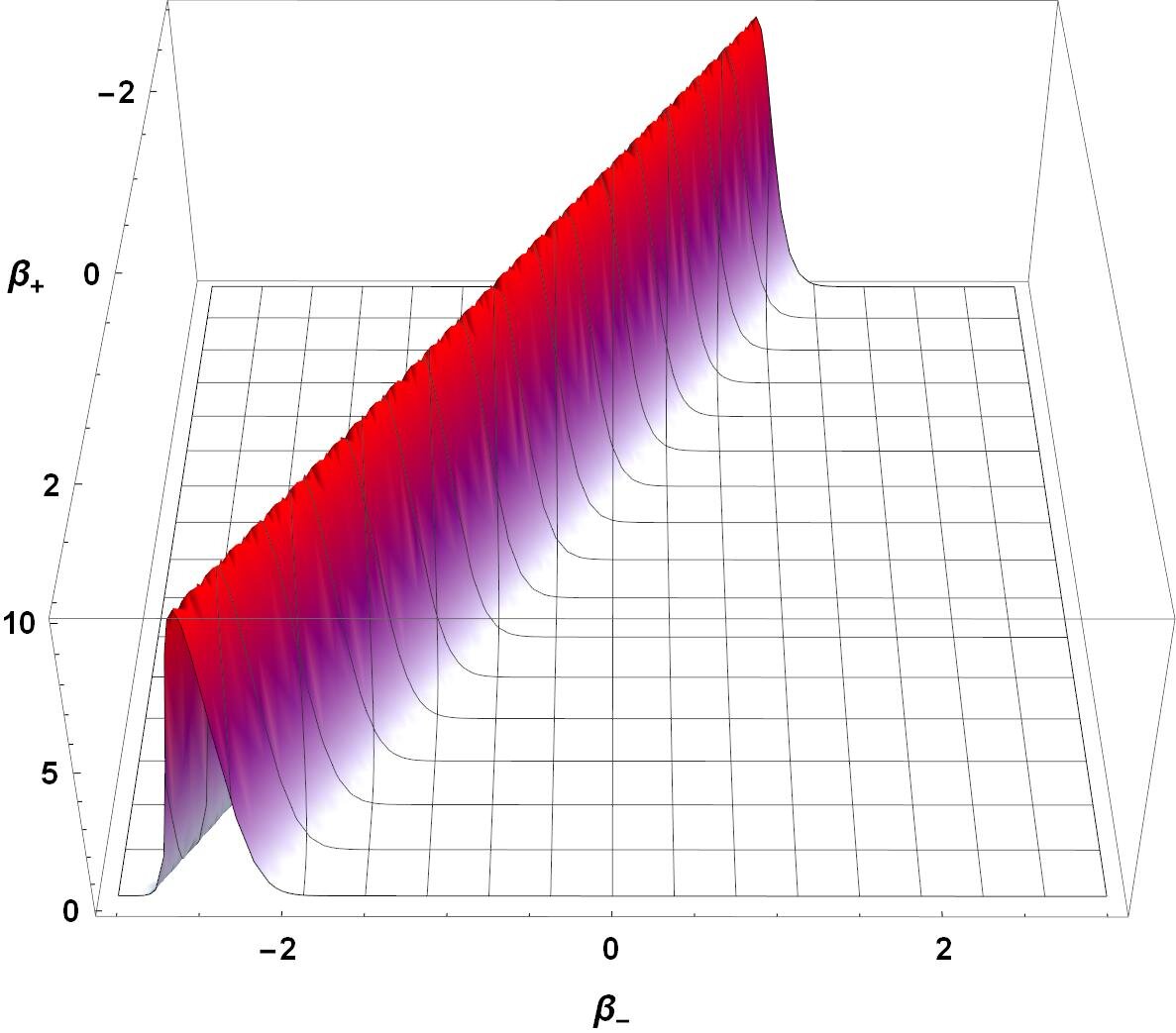}
\caption{$\alpha=-\frac{1}{4}$ $\hspace{1mm}$ $\Lambda=-1$ $\hspace{1mm}$ b=7 $\hspace{1mm}$ $\rho=0$}
  \label{2b}
\end{subfigure}
\begin{subfigure}{.4\textwidth}
  \centering
  \includegraphics[scale=.14]{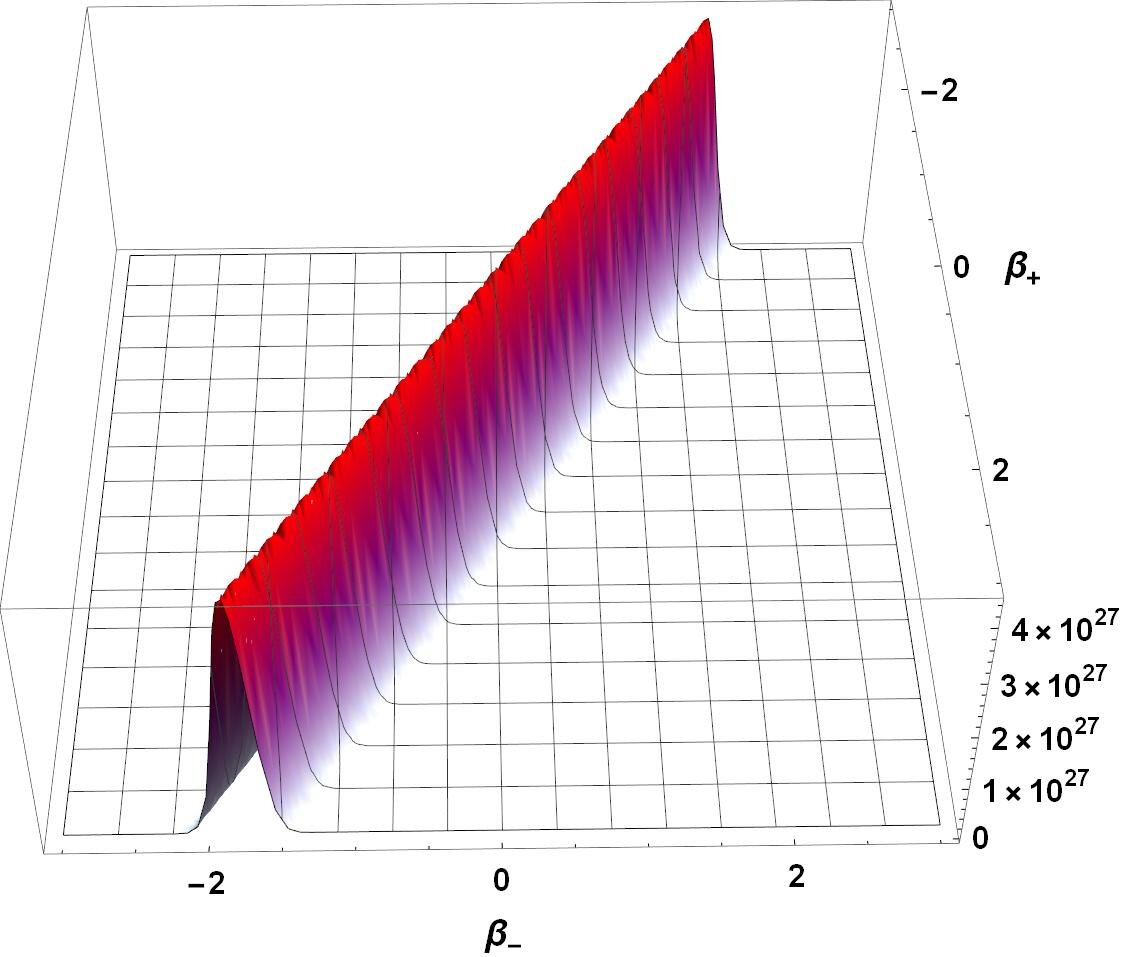}
\caption{$\alpha=\frac{1}{4}$ $\hspace{1mm}$ $\Lambda=-1$ $\hspace{1mm}$ b=0 $\hspace{1mm}$ $\rho=0$}
  \label{2c}
\end{subfigure}%
\begin{subfigure}{.4\textwidth}
  \centering
 \includegraphics[scale=.14]{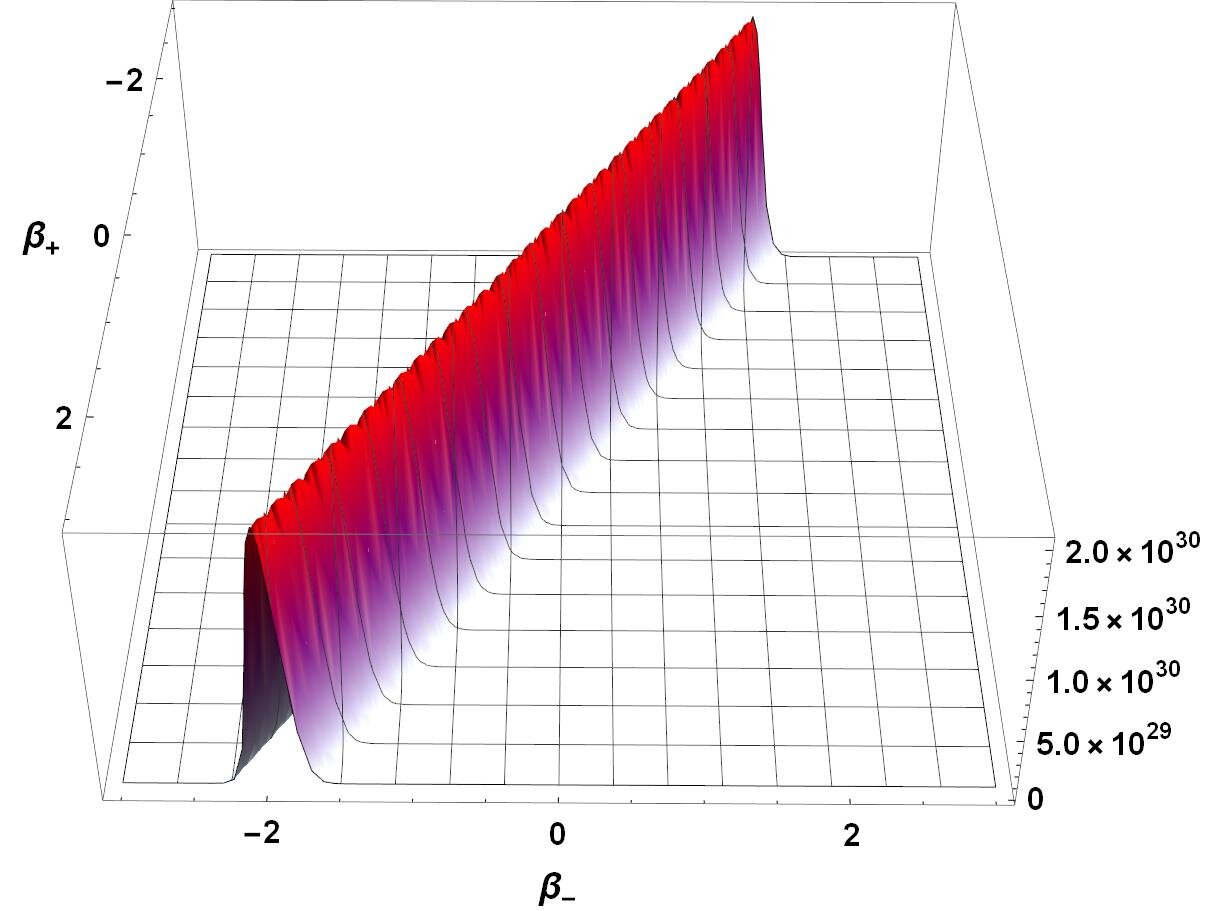}
\caption{$\alpha=\frac{1}{4}$ $\hspace{1mm}$ $\Lambda=-1$ $\hspace{1mm}$ b=7 $\hspace{1mm}$ $\rho=0$}
  \label{2d}
\end{subfigure}%
\caption{Plots of (\ref{79}) for $\abs{\Psi}^{2}$ when our aligned electromagnetic field and $\alpha$ are varied}
\label{fig 2}
\end{figure}

Using the following ansatz 

\begin{equation}\label{75}
\begin{aligned} 
\mathcal{S}^{4}_{(0)}=-6 \Lambda e^{4 \alpha-2 \sqrt{3} \beta_{-}-2  \beta_{+}}+\frac{1}{12}e^{2 \alpha+2 \sqrt{3} \beta_{-}+2 \beta_{+}}+
\alpha\text{x1}+\beta_{-} \text{x3}+\beta_{+}  \text{x2}
\end{aligned}
\end{equation}
the author found the following solutions to the Bianchi II Euclidean-signature Hamilton Jacobi equation corresponding to (\ref{7})

\begin{equation}\label{76}
\begin{aligned} 
&\left(\frac{\partial \mathcal{S}^{4}_{(0)}}{\partial \alpha}\right)^{2}-\left(\frac{\partial \mathcal{S}^{4}_{(0)}}{\partial \beta_{+}}\right)^{2}-\left(\frac{\partial \mathcal{S}^{4}_{(0)}}{\partial \beta_{-}}\right)^{2}+ \frac{1}{12} e^{4\alpha + 4\beta_+ +4\sqrt{3}\beta_-} +24\Lambda e^{6\alpha}+2b^{2}e^{2\alpha + 2\beta_+ +2\sqrt{3}\beta_-}+\rho=0 \\&
\mathcal{S}^{4}_{(0)  }= -6 \Lambda e^{4 \alpha-2 \sqrt{3} \beta_{-}-2 \beta_{+}}+\frac{1}{12} e^{2 \alpha+2 \sqrt{3} \beta_{-}+2 \beta_{+}}-2 \alpha
   \text{b}^2+\frac{1}{2} \beta_{-} \left(2 \sqrt{3}
   \text{b}^2+\sqrt{\rho}\right)+\frac{1}{2} \beta_{+} \left(2 \text{b}^2-\sqrt{3}
   \sqrt{\rho}\right).
\end{aligned}
\end{equation}
If we insert (\ref{76}) into (\ref{15}) we obtain 
\begin{equation}\label{77}
\begin{aligned} 
&-24 \Lambda e^{4 \alpha-2 \left(\sqrt{3} \beta_{-}+\beta_{+}\right)} \left(2 \frac{\partial \mathcal{S}_{(1)}}{\partial \alpha}+\sqrt{3}
   \frac{\partial \mathcal{S}_{(1)}}{\partial \beta_-}+\frac{\partial \mathcal{S}_{(1)}}{\partial \beta_+}+\text{B}\right) \\&+\frac{1}{6} e^{2 \left(\alpha+\sqrt{3} \beta_{-}+\beta_{+}\right)}
   \left(2 \frac{\partial \mathcal{S}_{(1)}}{\partial \alpha}-2 \sqrt{3} \frac{\partial \mathcal{S}_{(1)}}{\partial \beta_-}-2 \frac{\partial \mathcal{S}_{(1)}}{\partial \beta_+}+\text{B}+6\right) \\&-4 \text{b}^2
   \frac{\partial \mathcal{S}_{(1)}}{\partial \alpha}-2 \text{b}^2 \left(\sqrt{3}  \frac{\partial \mathcal{S}_{(1)}}{\partial \beta_-}+\frac{\partial \mathcal{S}_{(1)}}{\partial \beta_+}\right) \\&+\sqrt{\rho} \left(\sqrt{3}
   \frac{\partial \mathcal{S}_{(1)}}{\partial \beta_+}-\frac{\partial \mathcal{S}_{(1)}}{\partial \beta_-}\right)-2 \text{b}^2 \text{B}=0;
\end{aligned}
\end{equation}
which in accordance with our previous reasoning can be satisfied by the following simple solution 
\begin{equation}\label{78}
\begin{aligned} 
\mathcal{S}^{4}_{(1)}:=\frac{1}{2} \alpha (-\text{B}-2)+\frac{\sqrt{3} \beta_{-}}{2}+\frac{\beta_{+}}{2}.
\end{aligned}
\end{equation}
Inserting this into the source term of (\ref{16}) yields $-\frac{B^{2}}{2}$ which vanishes when $B=0$. Thus we have the following solution to the Bianchi II WDW equation when a cosmological constant, aligned electromagnetic field and stiff matter are present

\begin{equation}\label{79}
\begin{aligned} 
\Psi=e^{6 \Lambda e^{4 \alpha-2 \left(\sqrt{3} \beta_{-}+
\beta_{+}\right)}-\frac{1}{12} e^{2 \left(\alpha+\sqrt{3}
   \beta_{-}+\beta_{+}\right)}+2 \alpha \text{b}^2+\alpha-\frac{1}{2} \beta_{-} \left(2 \sqrt{3}
   \text{b}^2+\sqrt{\rho}+\sqrt{3}\right)+\frac{1}{2} \beta_{+} \left(-2
   \text{b}^2+\sqrt{3} \sqrt{\rho}-1\right)}.
\end{aligned}
\end{equation}
To understand what effects the aligned electromagnetic field ($b^{2}$) has on our wave function (\ref{79}) we construct a series of plots in \ref{fig 2}, and discuss them at the end of this paper.

We can obtain a solution for any Hartle-Hawking ordering parameter if we consider the case when only a cosmological constant is present. If we start with the following semi-classical term 
\begin{equation}\label{80}
\begin{aligned} 
\mathcal{S}^{5}_{(0)}=-6 \Lambda e^{4 \alpha-2 \sqrt{3} \beta_{-}-2  \beta_{+}}+\frac{1}{12}e^{2 \alpha+2 \sqrt{3} \beta_{-}+2 \beta_{+}}
\end{aligned}
\end{equation}
and insert it into (\ref{15}) we obtain the following $\mathcal{S}^{5}_{(1)}$
\begin{equation}\label{81}
\begin{aligned} 
\mathcal{S}^{5}_{(1)}:=\frac{1}{2} 
\alpha(-\text{B}-2)+
\beta_{-}\text{x1}+\beta_{+}\left(2-\sqrt{3} \text{x1}\right).
\end{aligned}
\end{equation}
When (\ref{81}) is inserted into (\ref{16}) we obtain this source term $-3 \text{B}^2-4 (1-2 \text{x1})^2$ which vanishes when $x1_{\pm}=\frac{1}{4} \left(2 \pm  \sqrt{3} \text{B}i\right)$. This allows us to construct two independent solutions to the Bianchi II WDW equation, one for each of the two possible values of $x1_{\pm}$ and sum them up to obtain
\begin{equation}\label{82}
\begin{aligned} 
\Psi=\left(e^{\frac{i \beta_{-}B}{2}}+e^{\frac{1}{2} i \sqrt{3} \beta_{+}
   \text{B}}\right) e^{ \left(\frac{1}{12} \left(72 \Lambda e^{4 \alpha-2
   \left(\sqrt{3} \beta_{-}+\beta_{+}\right)}-e^{2 \left(\alpha+\sqrt{3} \beta_{-}+\beta_{+}\right)}+6 \alpha
   (\text{B}+2)-3 i \text{B} \left(\beta_{-}+\sqrt{3} \beta_{+}\right)-6 \left(\sqrt{3}
   \beta_{-}+\beta_{+}\right)\right)\right)}.
\end{aligned}
\end{equation}

\begin{figure}
\centering
\begin{subfigure}{.4\textwidth}
  \centering
\includegraphics[scale=.18]{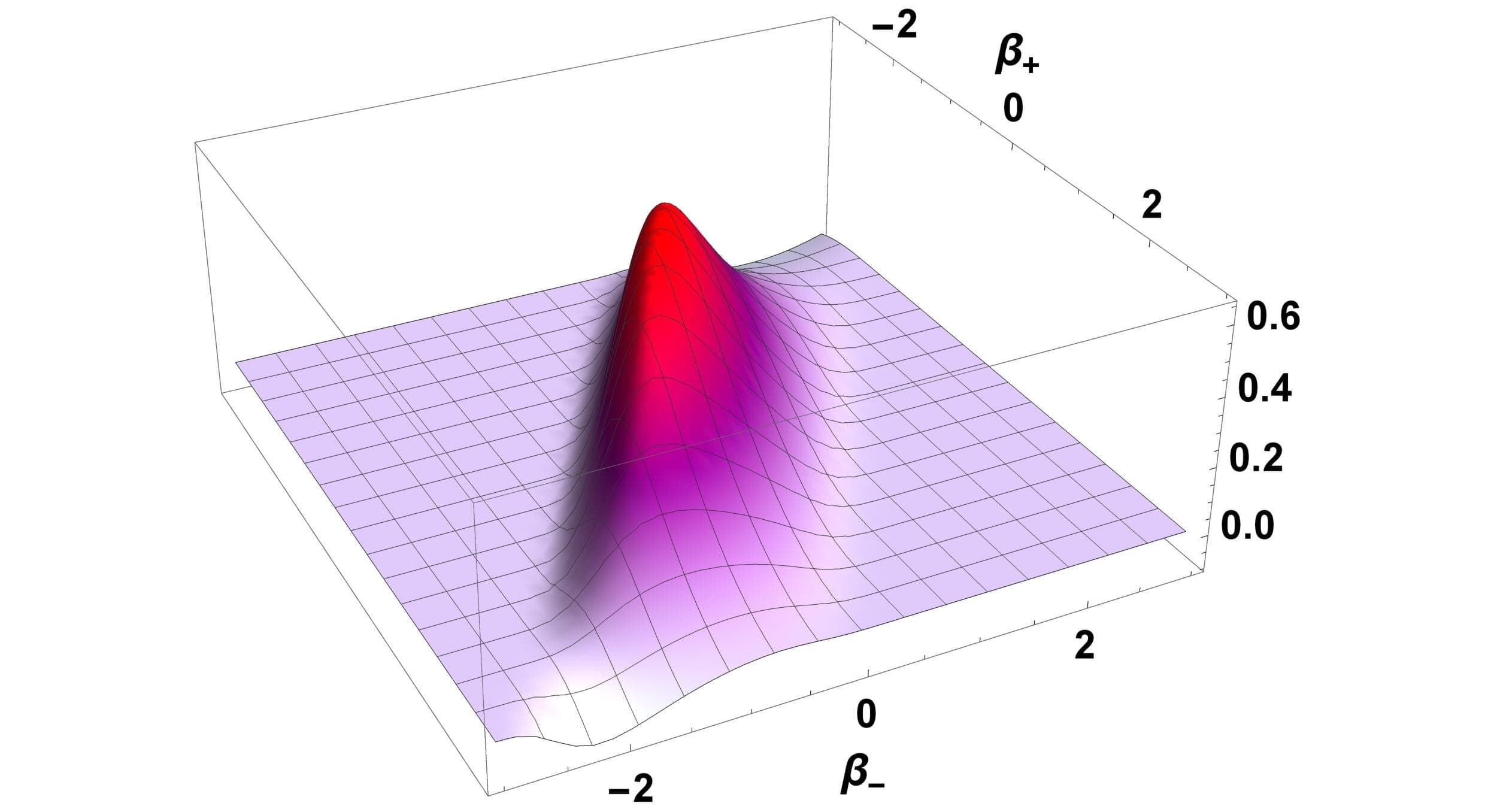}
\caption{$\alpha=-1$ $\hspace{1mm}$ $\Lambda=-1$ $\hspace{1mm}$ B=1}
  \label{3a}
\end{subfigure}%
\begin{subfigure}{.4\textwidth}
  \centering
\includegraphics[scale=.2]{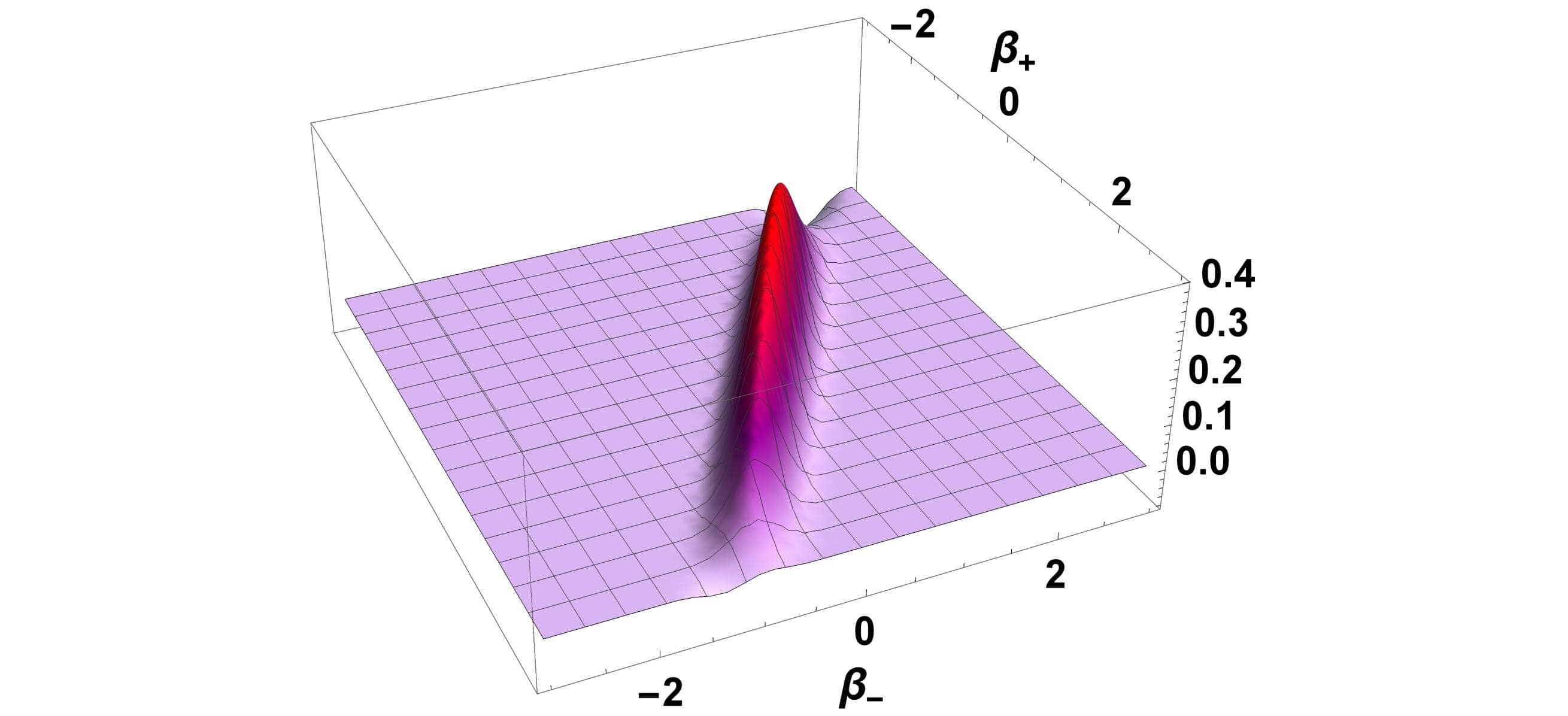}
\caption{$\alpha=0$ $\hspace{1mm}$ $\Lambda=-1$ $\hspace{1mm}$ B=1}
  \label{3b}
\end{subfigure}
\begin{subfigure}{.4\textwidth}
  \centering
  \includegraphics[scale=.18]{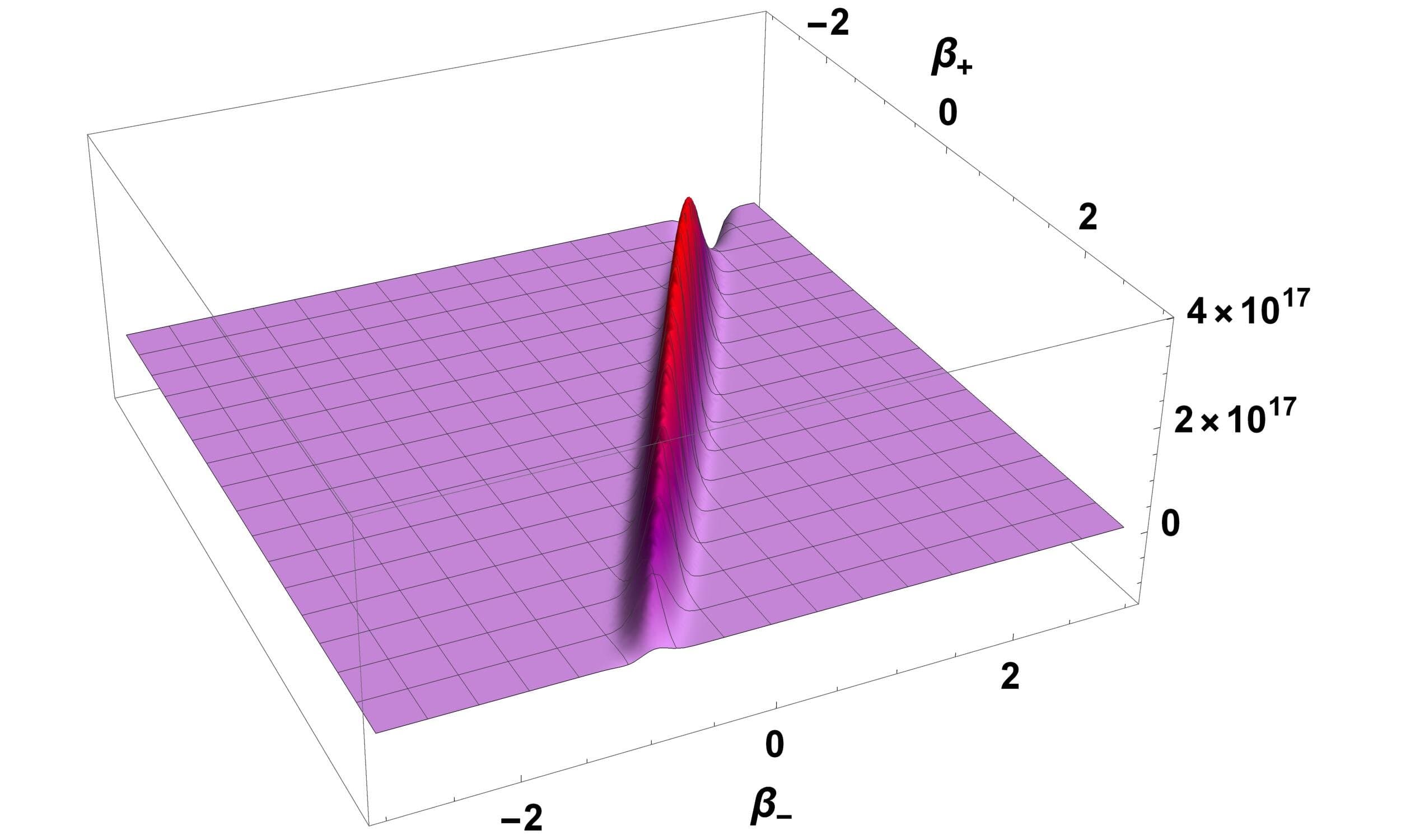}
\caption{$\alpha=\frac{1}{2}$ $\hspace{1mm}$ $\Lambda=-1$ $\hspace{1mm}$ B=1}
  \label{3c}
\end{subfigure}%
\caption{Plots of (\ref{82}) for $\abs{\Psi}^{2}$ for different values of $\alpha$.}
\label{fig 3}
\end{figure}

These solutions non-trivially depend on the ordering parameter. When $B=0$ these solutions are real, otherwise they are complex. We plot them for three different values of $\alpha$ in \ref{fig 3} and will discuss them in detail towards the end of this paper.

\section{\label{sec:level1}Closed Form 'Excited' States Of The $\Lambda \ne 0 $ Bianchi II Wheeler DeWitt Equation } 

The author was able to find the following $\phi_{0}$ for the case when only a cosmological constant is present.

\begin{equation}\label{83}
\begin{aligned} 
\phi^{5}_{0}:=\left(e^{\frac{1}{3} \left(3 \beta_+-\sqrt{3} \beta_-\right)}\right)^{m_{1}} \left(48 \Lambda e^{6 \alpha-4
   \sqrt{3} \beta_-}+e^{4 (\alpha+\beta_+)}\right)^{m_{2}}.
\end{aligned}
\end{equation}
This $\phi_{0}$ suggests that the 'excited' states of the quantum Bianchi II models when a cosmological constant is present have some interesting properties. When $\Lambda >0 $ our 'excited' states are scattering states because none of the terms exponentiated by our graviton excitation numbers $m_{1}$ and $m_{2}$ vanish for any real values of the Misner variables. However when $\Lambda < 0 $ the terms associated with $m_{1}$ don't vanish, while the term exponentiated by $m_{2}$ does vanish for real values of the Misner variables. Thus for $\Lambda < 0 $ our $m_{1}$ term can be any real number, while $m_{2}$ is restricted to being either zero or a positive integer. The excited' states for the quantum Bianchi II models when $\Lambda < 0 $ are hybrid scattering/bound states. This property is shared with the quantum Bianchi VIII models\cite{berkowitz2021bianchi}, which the author studied as well. The higher order $\phi^{5}_{k}$ terms in principle can be found by solving the rest of the transport equations. However, due to $\mathcal{S}^{5}_{(0) }$ possessing two terms with different $\alpha$ dependence, it is more difficult to solve these transport equations as opposed to the ones we encountered earlier. The author in trying to solve the $\phi^{5}_{1}$ terms computed an unenlightening integral expression which we will omit. The plots we will show will be leading order in $\phi$ 'excited' states. 

To construct our graphs we will set $m_{1}=-3$ and $m_{2}=1$ and graph the modulus squared of the following wave function

\begin{equation}\label{84}
\begin{aligned} 
\Psi_{Excited}=\sum_{m=1}^{10} \left(\phi^{5}_{0}\right)^{m}\Psi,
\end{aligned}
\end{equation}

where $\Psi$ is our exact wave function (\ref{82}). We will also graph  cross sections of our 3D plots when $\beta_{+}=0$.

\begin{figure}
\centering
\begin{subfigure}{.4\textwidth}
  \centering
 \includegraphics[scale=.16]{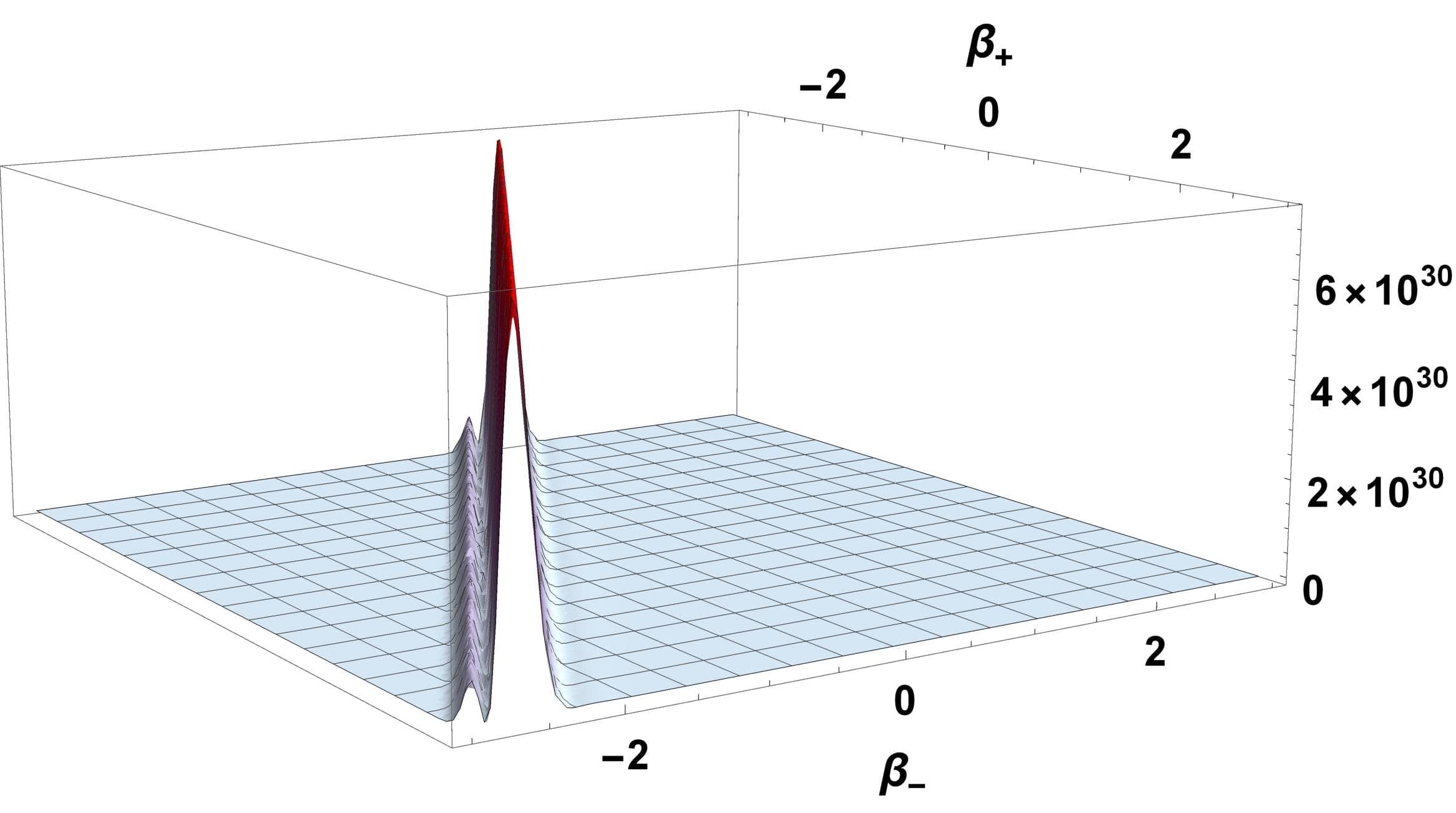}
\caption{$\alpha=-1$ $\hspace{1mm}$ $\Lambda=-1$ $\hspace{1mm}$ B=0}
  \label{4a}
\end{subfigure}%
\begin{subfigure}{.4\textwidth}
  \centering
 \includegraphics[scale=.16]{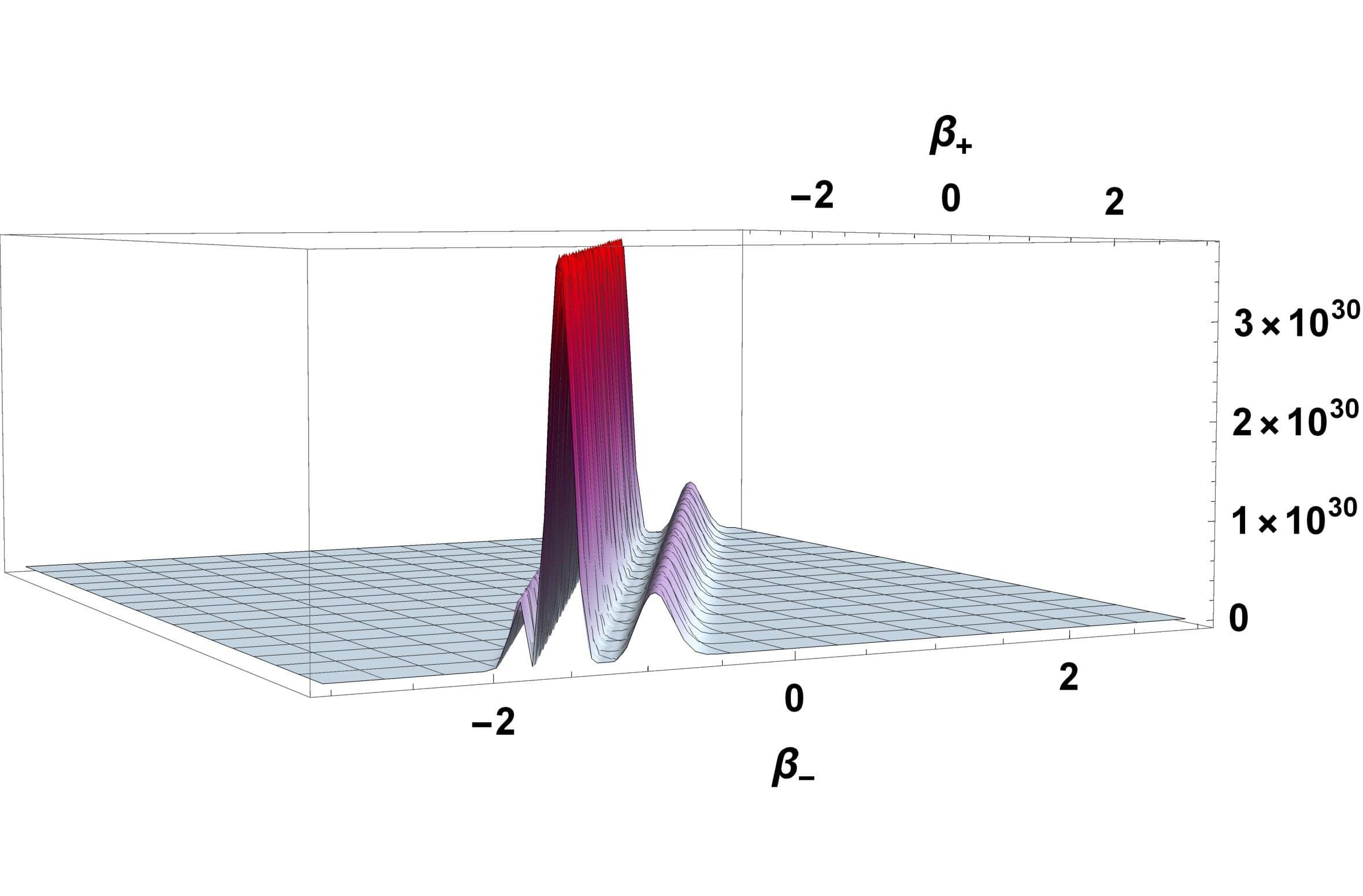}
\caption{$\alpha=0$ $\hspace{1mm}$ $\Lambda=-1$ $\hspace{1mm}$ B=0}
  \label{4b}
\end{subfigure}
\begin{subfigure}{.4\textwidth}
  \centering
 \includegraphics[scale=.18]{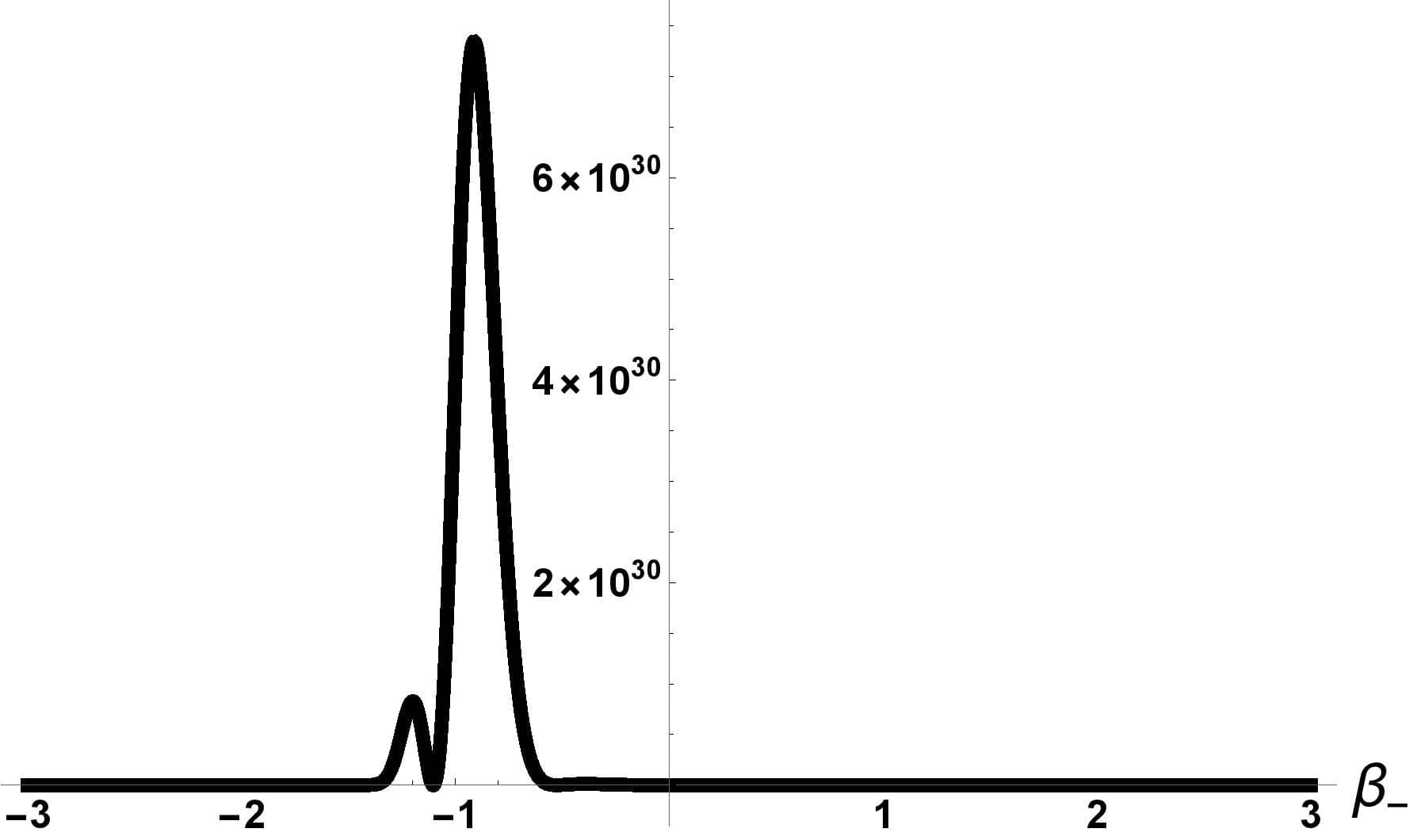}
\caption{$\alpha=-1$ $\hspace{1mm}$ $\Lambda=-1$ $\hspace{1mm}$ B=0}
  \label{4c}
\end{subfigure}%
\begin{subfigure}{.4\textwidth}
  \centering
 \includegraphics[scale=.18]{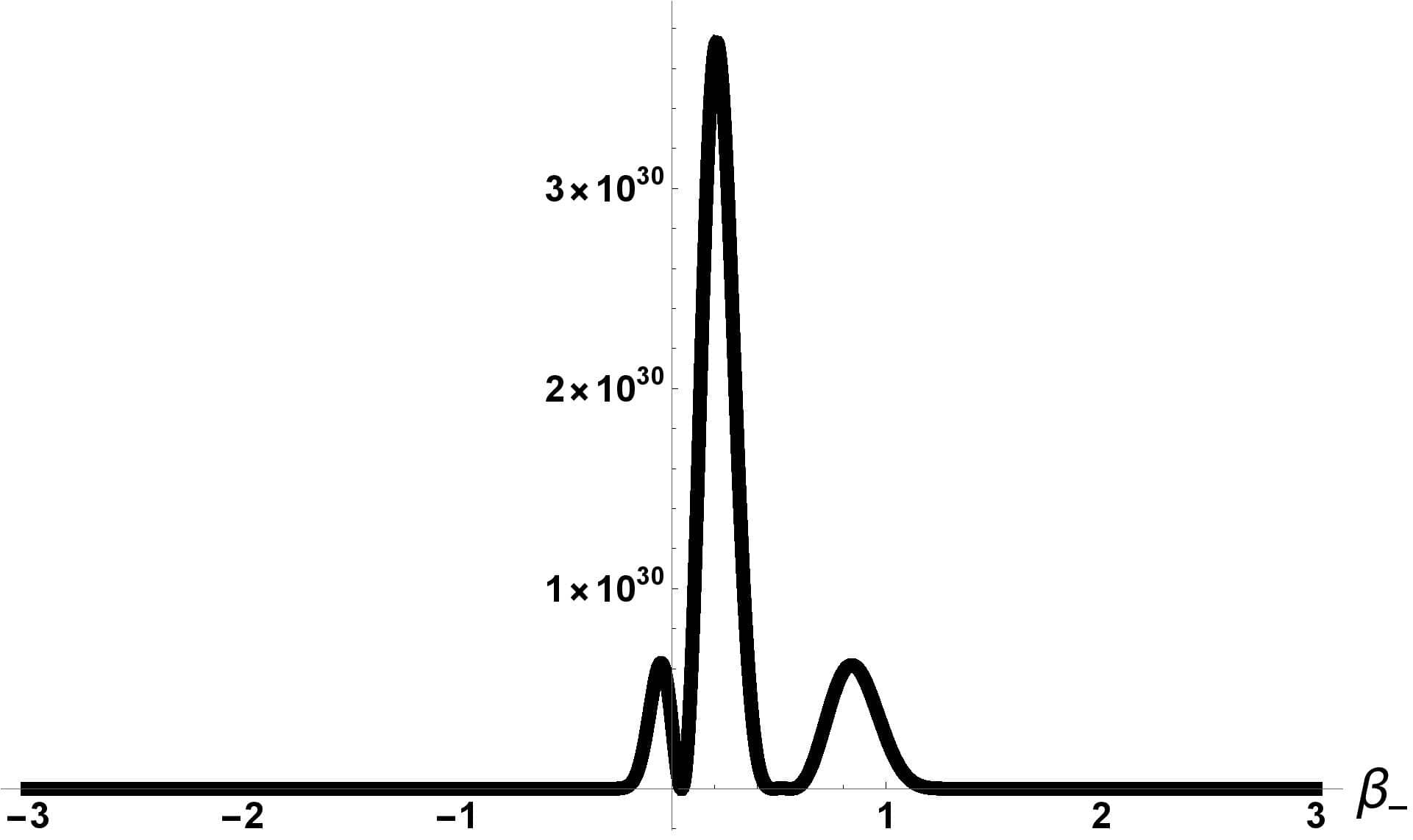}
\caption{$\alpha=0$ $\hspace{1mm}$ $\Lambda=-1$ $\hspace{1mm}$ B=0}
  \label{4d}
\end{subfigure}%
\caption{Plots of (\ref{84}) when $\abs{\Psi_{Excited}}^{2}$ for various values of $\alpha$ when $m_{1}=-3$ and $m_{2}=1$. }
\label{fig 4}
\end{figure}

\section{\label{sec:level1}Non-Commutative Quantum Bianchi II With Matter Sources} 

 In this section we shall study the quantum non-commutative Bianchi II models when an aligned electromagnetic field and stiff matter are present. To do so we shall use the following deformation of the ordinary commutation relations between the minisuperspace variables 

\begin{equation}\label{85}
\left[\alpha_{\mathrm{nc}}, \beta_{+\mathrm{nc}}\right]=\mathrm{i} \theta_{1}, \quad\left[\alpha_{\mathrm{nc}}, \beta_{-\mathrm{nc}}\right]=\mathrm{i} \theta_{2}, \quad\left[\beta_{-\mathrm{nc}}, \beta_{+\mathrm{nc}}\right]=\mathrm{i} \theta_{3}.
\end{equation}
This type of deformation of the configuration or phase space of a finite dimensional theory is employed in non-commutative quantum mechanics\cite{bertolami2005noncommutative,bertolami2006scaling,bastos2008weyl}. The non-commutative quantum Bianchi II models with stiff matter were thoroughly investigated in \cite{aguero2007noncommutative}. In this section we will follow their methodology and extend their results by including an aligned electromagnetic field.

The purpose of imposing these non-commutative relations on the minisuperspace of our Bianchi II models is to obtain a better understanding of how non-commutative space-time could have effected cosmological evolution in the early universe. Many theories of quantum gravity predict that space-time itself manifests some form of discretization. One way for this supposed discretization to manifest mathematically is in the coordinates$\left(t,x_{i}\right)$ of space-time possessing non vanishing commutation relations. For example in String Theory/M-Theory a non-commutative gauge theory emerges when describing the low energy excitations of open strings in the presence of a Neveu-Schwarz constant background B field\cite{douglas1998d,seiberg1999string}. As a result there has been a renewed interest in the study of non-commutative space-times.

A non-commutative space-time version of general relativity has been proposed\cite{garcia2003noncommutative} and in theory one can use it to directly study the full impact that non-commutative space-time has on classical cosmological evolution. However formulating general relativity in a non-commutative space-time results in a theory that is incredibly non-linear and very difficult to work with mathematically. A way to obtain some understanding of how non-commutative space-times can effect cosmology was proposed by \cite{garcia2002noncommutative}. Instead of directly studying the cosmology of a theory of gravity with a non-commutative space-time one can study a cosmology with deformed minisuperspace commutative relations as  presented in (\ref{85}). The justification for this can be summed up by saying that it is reasonable to expect that a full non-commutative space-time theory of gravity would result in some effects which can be captured by introducing non-commutativity in the minisuperspace of its homogeneous cosmologies. Thus by studying non-commutative minisuperspace homogeneous cosmologies we are studying an effective toy model of a non-commutative theory of gravity with its degrees of freedom reduced by imposing the symmetries present in homogeneous space-times. 

To begin the process of solving the non-commutative Bianchi II WDW equation we will implement the following Seiberg-Witten map\cite{seiberg1999string}

\begin{equation}\label{86}
\alpha_{\mathrm{nc}} \rightarrow \alpha-\frac{\theta_{1}}{2} p_{\beta_{+}}-\frac{\theta_{2}}{2} p_{\beta_{-}}, \quad \beta_{-\mathrm{nc}} \rightarrow \beta_{-}+\frac{\theta_{2}}{2} p_{\alpha}-\frac{\theta_{3}}{2} p_{\beta_{+}}, \quad \beta_{+\mathrm{nc}} \rightarrow \beta_{+}+\frac{\theta_{1}}{2} p_{\alpha}+\frac{\theta_{3}}{2} p_{\beta_{-}}.
\end{equation}
Doing so results in the following modified potential term for (\ref{7}) when $\Lambda=0$

\begin{equation}\label{87}
\begin{aligned}
\mathrm{U}\left(\Omega, \beta_{\pm}\right)=&\frac{1}{12} \mathrm{e}^{4\left[\alpha+\beta_{+}+\sqrt{3} \beta_{-}-\frac{\mathrm{i} \theta_{1}}{2}\left(\frac{\partial}{\partial \alpha}-\frac{\partial}{\partial \beta_{+}}\right)-\frac{\mathrm{i} \theta_{2}}{2}\left(\sqrt{3} \frac{\partial}{\partial \alpha}-\frac{\partial}{\partial \beta_{-}}\right)+\frac{\mathrm{i} \theta_{3}}{2}\left(\sqrt{3} \frac{\partial}{\partial \beta_{+}}-\frac{\partial}{\partial \beta_{-}}\right)\right]} \\& +2b^{2} \mathrm{e}^{2\left[\alpha+\beta_{+}+\sqrt{3} \beta_{-}-\frac{\mathrm{i} \theta_{1}}{2}\left(\frac{\partial}{\partial \alpha}-\frac{\partial}{\partial \beta_{+}}\right)-\frac{\mathrm{i} \theta_{2}}{2}\left(\sqrt{3} \frac{\partial}{\partial \alpha}-\frac{\partial}{\partial \beta_{-}}\right)+\frac{\mathrm{i} \theta_{3}}{2}\left(\sqrt{3} \frac{\partial}{\partial \beta_{+}}-\frac{\partial}{\partial \beta_{-}}\right)\right]} +\rho.
\end{aligned}
\end{equation}
After applying the following coordinate transformation 

\begin{equation}\label{88}
\xi=\Omega+\beta_{+}+\sqrt{3} \beta_{-}, \quad \kappa=\Omega+\frac{\sqrt{3}}{3} \beta_{-}, \quad \lambda=\Omega-2 \beta_{+}+\sqrt{3} \beta_{-}
\end{equation}
and applying the generalized Baker-Campbell-Hausdorff formula
\begin{equation}\label{89}
\mathrm{e}^{\eta(\hat{\mathrm{A}}+\hat{\mathrm{B}})}=\mathrm{e}^{-\eta^{2}[\hat{\mathrm{A}}, \hat{\mathrm{B}}]} \mathrm{e}^{\eta \hat{\mathrm{A}}} \mathrm{e}^{\eta \hat{\mathrm{B}}}
\end{equation}
we obtain the following WDW equation 

\begin{equation}\label{90}
\begin{aligned}
& -B \frac{\partial \Psi}{\partial \xi}-3 \frac{\partial^{2} \Psi}{\partial \xi^{2}} 
+\frac{2}{3} \frac{\partial^{2} \Psi}{\partial \kappa^{2}}-B \frac{\partial \Psi}{\partial \kappa}-6 \frac{\partial^{2} \Psi}{\partial \lambda^{2}}-B\frac{\partial \Psi}{\partial \lambda} \\& + \frac{1}{12} \mathrm{e}^{4 \xi} \mathrm{e}^{-2 \mathrm{i} \theta_{1} \frac{\partial}{\partial \kappa}} \mathrm{e}^{-6 \mathrm{i} \theta_{1} \frac{\partial}{\partial \lambda}} \mathrm{e}^{\frac{-4 \sqrt{3}}{3} \mathrm{i} \theta_{2} \frac{\partial}{\partial \kappa}} \mathrm{e}^{-\frac{2 \sqrt{3}}{3} \mathrm{i} \theta_{3} \frac{\partial}{\partial \kappa}} \mathrm{e}^{-6 \sqrt{3} \mathrm{i} \theta_{3} \frac{\partial}{\partial \lambda}}\Psi \\&+2b^{2} \mathrm{e}^{2 \xi} \mathrm{e}^{- \mathrm{i} \theta_{1} \frac{\partial}{\partial \kappa}} \mathrm{e}^{-3 \mathrm{i} \theta_{1} \frac{\partial}{\partial \lambda}} \mathrm{e}^{\frac{-2\sqrt{3}}{3} \mathrm{i} \theta_{2} \frac{\partial}{\partial \kappa}} \mathrm{e}^{-\frac{\sqrt{3}}{3} \mathrm{i} \theta_{3} \frac{\partial}{\partial \kappa}} \mathrm{e}^{-3 \sqrt{3} \mathrm{i} \theta_{3} \frac{\partial}{\partial \lambda}}\Psi+\rho\Psi=0.
\end{aligned}
\end{equation}

To solve this equation we will insert this ansatz into it,   $\Psi=f(\xi)e^{\mathrm{i}c1\kappa}e^{\mathrm{i}c2\lambda}$ where c1 and c2 are constants while keeping in mind that $\mathrm{e}^{\mathrm{i} \theta \frac{\partial}{\partial \mathrm{x}} \mathrm{e}^{\eta \mathrm{x}}} \equiv \mathrm{e}^{\mathrm{i} \eta \theta} \mathrm{e}^{\eta \mathrm{x}}$, resulting in 

\begin{equation}\label{91}
\begin{aligned}
&f(\xi) \left(24 \text{b}^2 e^{2 e+2 w}+e^{4 e+4 w}+g\right)-12 \left(\text{B}
   \frac{\partial f}{\partial \xi}+3 \frac{\partial^{2} f}{\partial \xi^{2}}\right)=0 \\& w=\frac{1}{2} \left(\theta_{1} (\text{c1}+3 \text{c2})+\frac{2 \theta_{2}
   \text{c1}+\theta_{3}\text{c1}+9 \theta_{3} \text{c2}}{\sqrt{3}}\right) \\&g=-12 i \text{B} \text{c1}-12 i \text{B} \text{c2}+12 \rho-8 \text{c1}^2+72
   \text{c2}^2.
\end{aligned}
\end{equation}
The solution to this equation (\ref{91}) is the following 

\begin{equation}\label{92}
\begin{aligned}
f(\xi)=e^{\frac{1}{12} \left(-2 \text{B} (\xi+w)-e^{2 (\xi+w)}\right)} \left(e^{2
   (\xi+w)}\right)^{\frac{\sqrt{B^2+g}}{12}}
   U\left(\text{b}^2+\frac{1}{12}
   \left(\sqrt{\text{B}^2+g}+6\right),\frac{1}{6}
   \left(\sqrt{\text{B}^2+g}+6\right),\frac{1}{6} e^{2 (\xi+w)}\right)
\end{aligned}
\end{equation}
where U is the hypergeometric U function. There is a generalized Laguerre polynomial which also satisfies (\ref{91}) but it yields solutions which do not appear to be physical. Using our ansatz, (\ref{88}), and (\ref{92}) one can express the solutions for the non-commutative Bianchi II WDW equation (\ref{7} \ref{87}). In what follows we will set $\theta_{2}=\theta_{1}, \quad \theta_{3}=\theta_{1}, \quad \rho=0$. We will present a series of plots for our non-commutative wave $\abs{\int^{\infty}_{-\infty}e^{-1.5(c1-1.3)^{2}}e^{ic1\kappa}f(\xi)dc1}^{2}$ function and discuss them at the end of this paper.

\begin{figure}
\centering
\begin{subfigure}{.4\textwidth}
  \centering
\includegraphics[scale=.13]{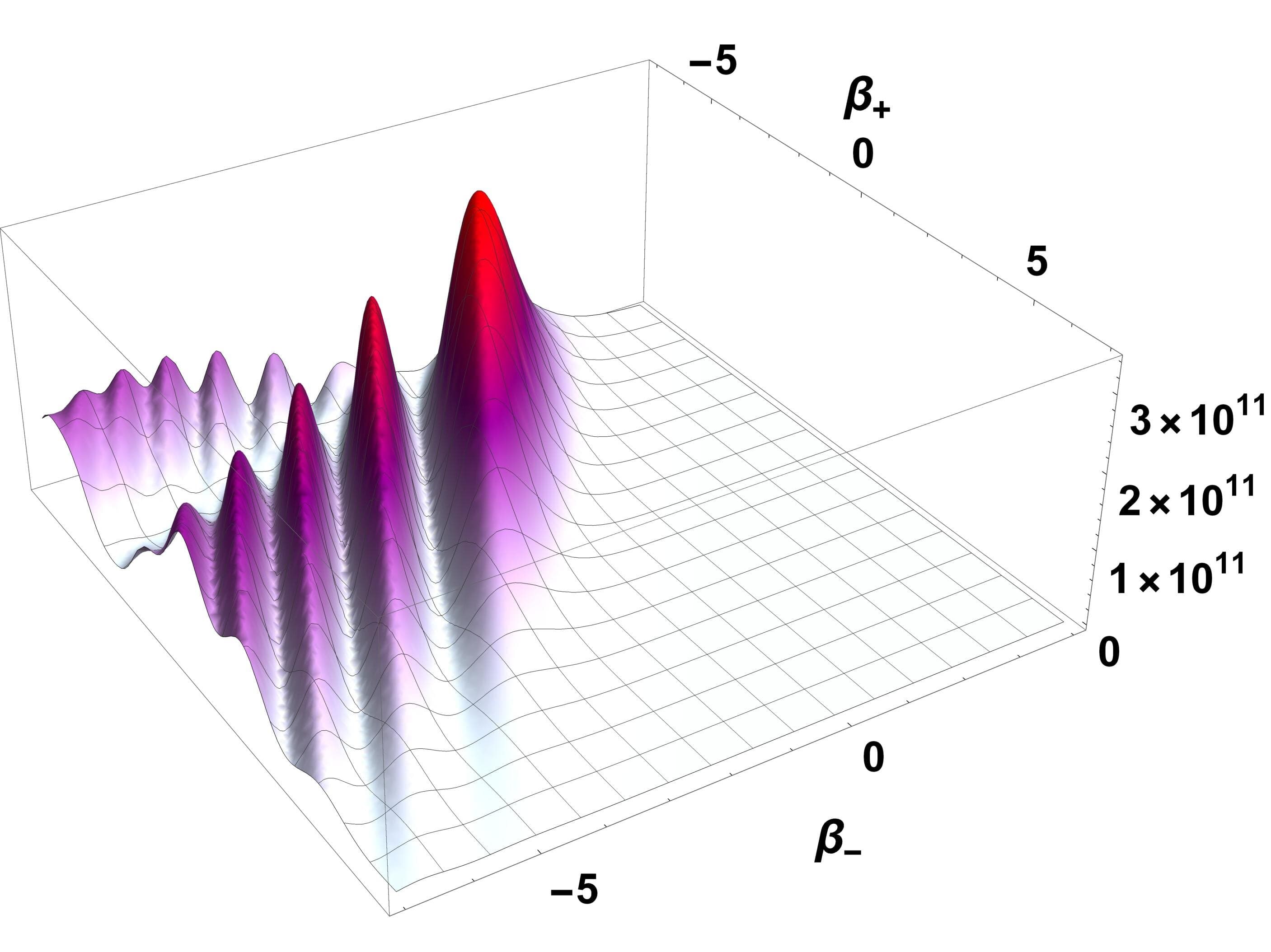}
\caption{$\alpha=-2$ $\hspace{1mm}$ $\theta_{1}=1.5$ $\hspace{1mm}$ $B=0$ $\hspace{1mm}$ $b=0$}
  \label{5a}
\end{subfigure}
\begin{subfigure}{.4\textwidth}
  \centering
\includegraphics[scale=.14]{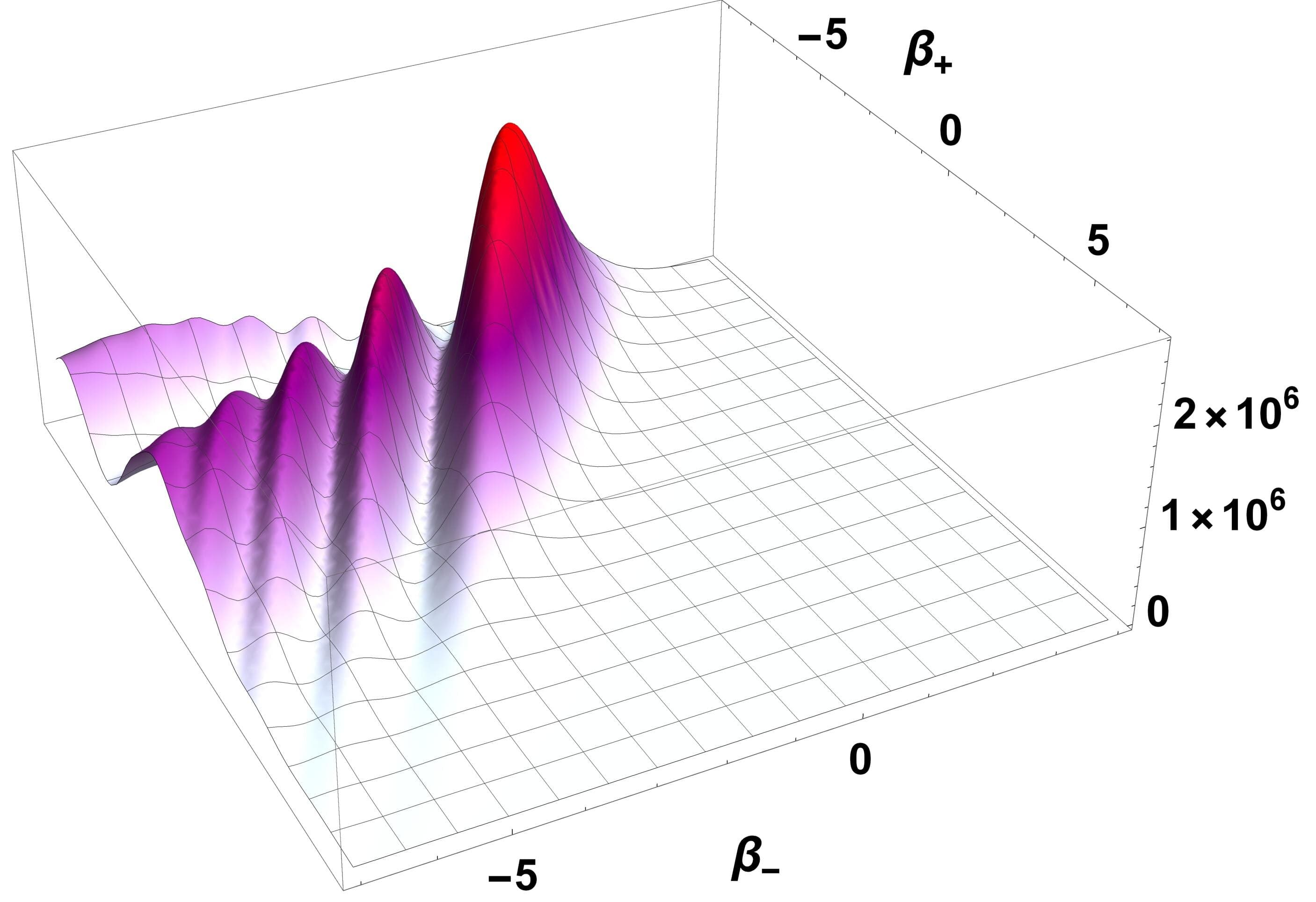}
\caption{$\alpha=-2$ $\hspace{1mm}$ $\theta_{1}=1.5$ $\hspace{1mm}$ $B=0$ $\hspace{1mm}$ $b=4$}
  \label{5b}
\end{subfigure}%
\caption{These plots of $\abs{\int^{\infty}_{-\infty}e^{-1.5(c1-1.3)^{2}}e^{ic1\kappa}f(\xi)dc1}^{2}$ represent a small subset of the wave functions we can obtain if we manipulate the non-commutative parameters $\theta_{i}$. }
\label{fig 5}
\end{figure}

\section{\label{sec:level1}Quantum Vacuum Bianchi VII$_{h=0}$ 'Ground' And 'Excited' States} 

Moving on to the Bianchi VII$_{h=0}$ models, these two solutions were found for its Euclidean-Signature Hamilton-Jacobi equation(6, 14) by \cite{obregon1996psi}

\begin{equation}\label{93}
\begin{aligned}
\mathcal{S}^{6}_{(0) } =\frac{1}{3} e^{2 (\alpha+\beta_{+})} \cosh \left(2 \sqrt{3} \beta_{-}\right)
\end{aligned}
\end{equation}

\begin{equation}\label{94}
\begin{aligned}
\mathcal{S}^{7}_{(0) } =\frac{1}{3} e^{2 (\alpha+\beta_{+})} \cosh \left(2 \sqrt{3} \beta_{-}\right)+\text{x1}
   e^{2 \alpha+2 \beta_{+}}
\end{aligned}
\end{equation}
where x1 is an arbitrary constant. 

Starting with (\ref{93}) if we insert it into (\ref{15}) and employ the methodology that we used to solve the Bianchi II $\mathcal{S}_{(1) }$ equation (\ref{27}) we obtain 

\begin{equation}\label{95}
\begin{aligned}
\mathcal{S}^{6}_{(1) } =\alpha \text{x1}+\frac{1}{12} \log \left(\sinh \left(2 \sqrt{3} \beta_{-}\right)\right)
   (\text{B}+2 \text{x1}-2 \text{x2}+6)+\beta_{+} \text{x2}
\end{aligned}
\end{equation}
where both x1 and x2 are arbitrary constants. Inserting this term into the source term of equation (\ref{16}) results in an expression which vanishes when our arbitrary constants equals

\begin{equation}\label{96}
\begin{aligned}
\left\{\text{x1}= \frac{1}{24} \left(-\text{B}^2-12
   \text{B}-36\right),\text{x2}= \frac{1}{24}
   \left(36-\text{B}^2\right)\right\}.
\end{aligned}
\end{equation}
This allow us to write down the following solution to the Bianchi VII$_{h=0}$ WDW equation

\begin{equation}\label{97}
\begin{aligned}
\Psi=e^{\left(\frac{1}{24} (\text{B}+6) (a (\text{B}+6)+\beta_{+}
   (\text{B}-6))-\frac{1}{3} e^{2 (\alpha+\beta_{+})} \cosh \left(2 \sqrt{3}
   \beta_{-}\right)\right)}.
\end{aligned}
\end{equation}
This solution was first reported by \cite{obregon1996psi}.

For the Bianchi VII$_{h=0}$ 'exicted' states the author found the following solutions to the $\phi_{0}$ equations for (\ref{93})

\begin{equation}\label{98}
\begin{aligned} 
\phi^{6}_{0}:=e^{6 m_{2}\alpha-6 m_{1}\beta_{+} } \sinh\left(2 \sqrt{3} \beta_{-}\right)^{m_{1}+m_{2}}.
\end{aligned}
\end{equation}
Due to $\sinh\left(2\sqrt{3}\beta_{-}\right)$ vanishing we must restrict the values of $m_{1}$ and $m_{2}$ so that their sum $m_{1}+m_{2}$ always equals a positive integer or zero. Using (\ref{98}) we can construct semi-classical 'excited' states. We can find closed form solutions to the Bianchi VII$_{h=0}$ WDW equation by inserting $\phi^{6}_{0}\Psi$ into it, where we used (\ref{97}) for our $\Psi$. By doing so we will find that it is satisfied when $m_{1}= \frac{1}{216} \left(\text{B}^2+84\right)$ and $m_{2}=
   \frac{1}{216} \left(132-\text{B}^2\right)$. This leads to the following closed form solution 

\begin{equation}\label{99}
\begin{aligned}
&\Psi_{excited}=\sinh \left(2 \sqrt{3} \beta_{-}\right) e^{ \left(\frac{1}{72} \left(-24 e^{2 (\alpha+\beta_{+})}
   \cosh \left(2 \sqrt{3} \beta_{-}\right)+\alpha (\text{B} (\text{B}+36)+372)+\beta_{+}
   \left(\text{B}^2-276\right)\right)\right)}.
\end{aligned}
\end{equation}

This solution has a strange property though. The potential for the Bianchi VII$_{h=0}$ models is invariant under the reflection of $\beta_{-}\to -\beta_{-}$ while our 'excited state isn't. This is another instance\cite{berkowitz2019new} in which the WDW equation admits solutions which do not respect the symmetry of its potential. However $\abs{\Psi_{excited}}^{2}$ does preserve the symmetry of the potential. This is important because if all of our observables are dependent upon $\abs{\Psi_{excited}}^{2}$, then the symmetry which is broken by (\ref{99}) may not bear any practical consequences. Nonetheless the physical implications of symmetry breaking solutions of the Wheeler DeWitt equation is a topic which deserves to be investigated more. 

Furthermore it is difficult to obtain an 'excited' state wave function which behaves in a way which is easy to interpret qualitatively. Thus we had to pick a specific value of the operator ordering parameter $B \geq 23$ to obtain solutions which are intuitive to interpret. The difficultly in computing 'excited' states for the vacuum Bianchi II and VII$_{h=0}$ in comparison to the Bianchi IX\cite{bae2015mixmaster} and Taub models\cite{berkowitz2020towards} may be a result of topological differences between their space-times or in mathematical difference of the Bianchi Lie algebras which define them.

\begin{figure}
\centering
\begin{subfigure}{.4\textwidth}
  \centering
\includegraphics[scale=.13]{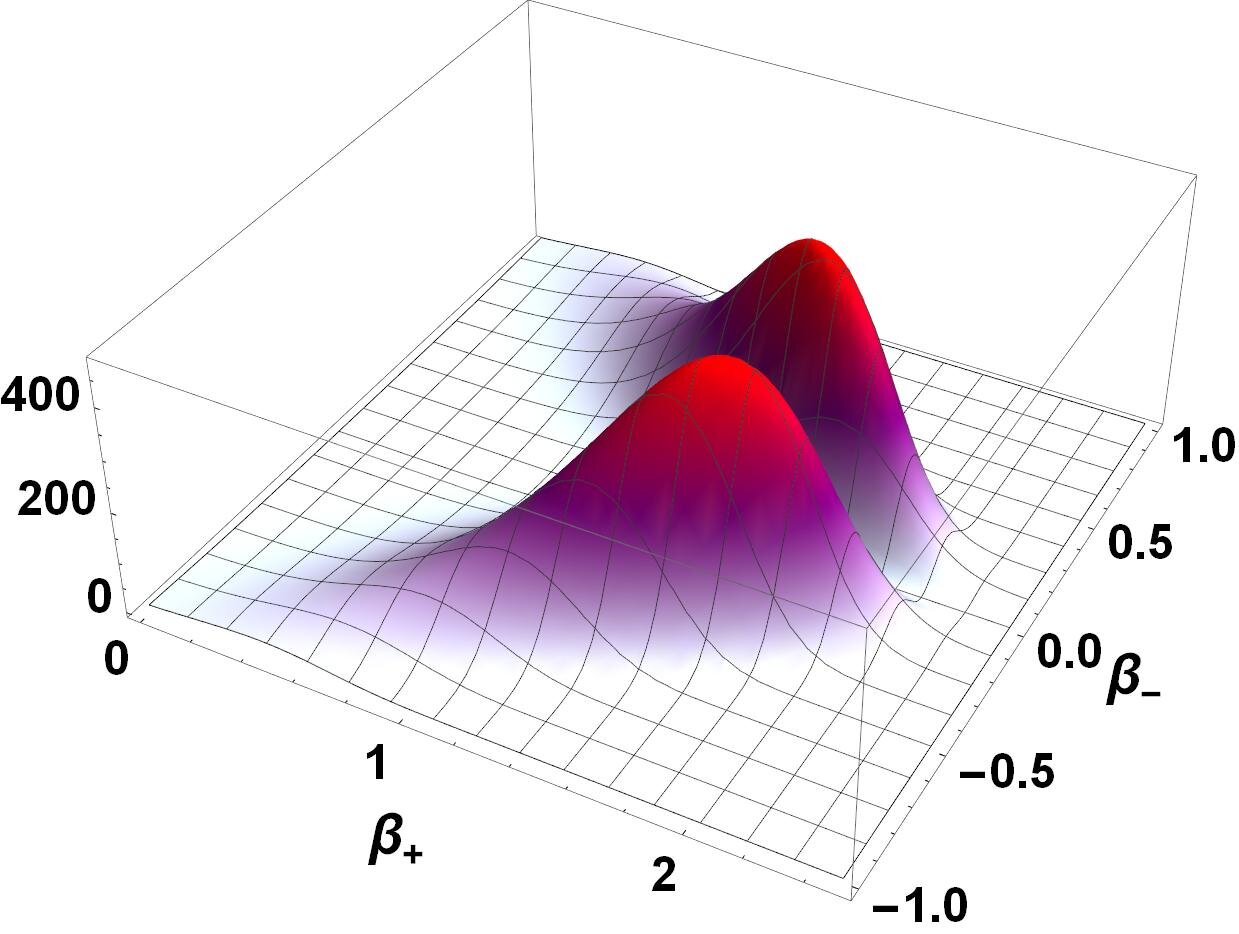}
\caption{$\alpha=-1$ $\hspace{1mm}$ B=23}
  \label{6a}
\end{subfigure}
\begin{subfigure}{.4\textwidth}
  \centering
\includegraphics[scale=.16]{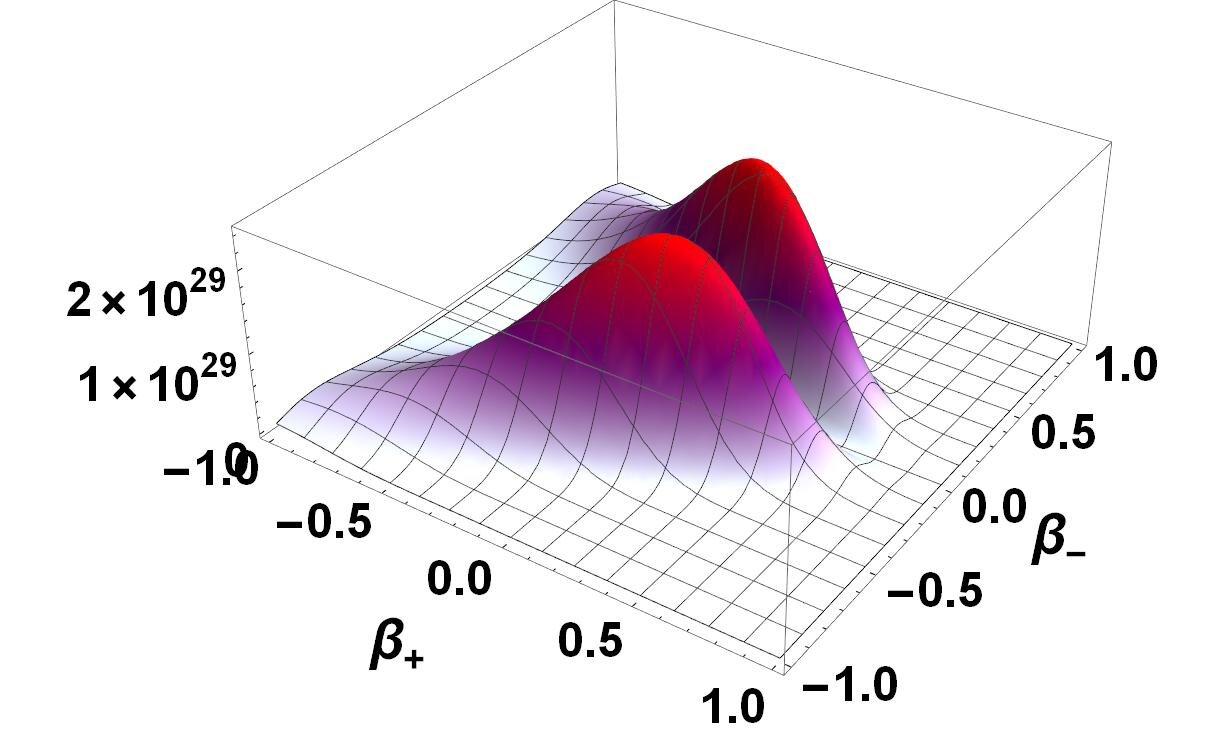}
\caption{$\alpha=\frac{1}{2}$ $\hspace{1mm}$ B=23}
  \label{6b}
\end{subfigure}%
\caption{Plots of (\ref{99}) for $\abs{\Psi_{excited}}^{2}$ for various values of $\alpha$ for a specific value of the ordering parameter.}
\label{fig 6}
\end{figure}

\section{\label{sec:level1}Quantum Vacuum Bianchi VII$_{h=0}$ 'Ground' States With Matter Sources} 

For the case when only an aligned electromagnetic field and stiff matter are present the author found the following solution to the Euclidean-signature Hamiliton Jacobi equation corresponding to (\ref{7}) 

\begin{equation}\label{100}
\begin{aligned}
\mathcal{S}^{7}_{(0)} =\frac{1}{6} \left(3 \text{b}^2 \left(-2 \alpha+\sqrt{3} \beta_{-}+\beta_{+}\right)+2 e^{2 (\alpha+\beta_{+})}
   \cosh \left(2 \sqrt{3} \beta_{-}\right)+\frac{2 \rho (\alpha+\beta_{+})}{\text{b}^2}\right).
\end{aligned}
\end{equation}
Inserting this expression in (\ref{15}) and seeking an $\mathcal{S}_{(1)}$ which is linear in the Misner variables results in 

\begin{equation}\label{101}
\begin{aligned}
\mathcal{S}^{7}_{(1)} =\frac{2 \beta_{+} \left(3 \text{b}^4-\rho\right)}{3 \text{b}^4}-\frac{\alpha \left(3
   \text{b}^4 \text{B}+6 \text{b}^4+4 \rho\right)}{6 \text{b}^4}.
\end{aligned}
\end{equation}
If we insert (\ref{101}) into the source term of (\ref{16}) we obtain $\frac{8 \rho}{b^4}-\frac{B^2}{2}-6=0$ which vanishes when our ordering parameter equals $B=\frac{2 \sqrt{4 \rho-3 b^4}}{b^2}$. This allow us to write down the following solution which satisfies (\ref{7}) when $\Lambda=0$ and $B=\frac{2 \sqrt{4 \rho-3 b^4}}{b^2}$

\begin{equation}\label{102}
\begin{aligned}
\Psi=e^{ \left(\frac{1}{6} \left(-3 b^2 \left(-2 \alpha+\sqrt{3} \beta_{-}+\beta_{+}\right)-2
   e^{2 (\alpha+\beta_{+})} \cosh \left(2 \sqrt{3} \beta_{-}\right)-\frac{2 \rho
   (\alpha+\beta_{+})}{b^2}\right)+\frac{2 \rho (\alpha+\beta_{+})}{3 b^4}+\frac{\alpha
   \sqrt{4 \rho-3b^4}}{b^2}+\alpha-2 \beta_{+}\right)}.
\end{aligned}
\end{equation}
In order for our ordering parameter to be real we require that $\rho \geq \frac{3}{4}b^{4}$. Nonetheless if one were interested in studying an asymptotic Bianchi VII$_{h=0}$ wave function, then they may be content to only include the first two terms of the expansion which we have calculated, or they can solve for higher order terms using our transport equations. 

To accommodate a cosmological constant we have to make our stiff matter term $\rho=3b^4$, which results in the following $\mathcal{S}_{(0)}$

\begin{equation}\label{103}
\begin{aligned}
\mathcal{S}^{8}_{(0)}=-3 \Lambda e^{4 \alpha+2 \sqrt{3} \beta_{-}-2 \beta_{+}}+\frac{1}{3} e^{2 (\alpha+\beta_{+})} \cosh \left(2 \sqrt{3}
   \beta_{-}\right)+\frac{1}{2} \text{b}^2 \left(\sqrt{3} \beta_{-}+3 \beta_{+}\right).
\end{aligned}
\end{equation}
Using the alternative operator ordering presented in \cite{giampieri1991new} we can satisfy the Bianchi VII$_{h=0}$ WDW equation when a cosmological constant, aligned electromagnetic field and stiff matter are present using just (\ref{103}). However for our purposes we will only consider semi classical wave functions $\psi=e^{-\mathcal{S}_{(0)}}$ of the form dictated by (\ref{7}). If we only consider the Bianchi VII$_{h=0}$ models with a cosmological constant our $\mathcal{S}_{(0)}$ term simplifies to  

\begin{equation}\label{104}
\begin{aligned}
\mathcal{S}^{9}_{(0)}=\frac{1}{3} e^{2 (\alpha+\beta_{+})} \cosh \left(2 \sqrt{3} \beta_{-}\right)-3 \Lambda e^{4 \alpha\pm_{2}
   \sqrt{3} \beta_{-}-2 \beta_{+}}.
\end{aligned}
\end{equation}
The $\pm$ signs of the $\beta_{-}$ term in (\ref{104}) is a result of the fact that the Bianchi VII$_{h=0}$ potential when only a cosmological constant is present is invariant under reflection of $\beta_{-}\to\beta_{-}$. Via (\ref{15}) and (\ref{104}) we find the following $\mathcal{S}_{(1)}$ term 

\begin{equation}\label{105}
\begin{aligned}
\mathcal{S}^{9}_{(1)}=\frac{1}{2} \alpha (-\text{B}-2)+2 \beta_{+}.
\end{aligned}
\end{equation}
Using the methodology that we have presented throughout this paper we can show that a closed form solution to the WDW equation exists when our ordering parameter equals $B= \pm 2  \sqrt{3}i$ using just (\ref{16}), (\ref{104}), and (\ref{105}). 

In order to determine how our aligned electromagnetic field effects our quantum Bianchi VII$_{h=0}$ wave functions we will compare the semi classical wave functions constructed from (\ref{103}) and (\ref{104}). We can group average over the reflection symmetry $\beta_{-}\to{-\beta_{-}}$
 present in the  Bianchi VII$_{h=0}$ potential when only $\Lambda$ is present to obtain

\begin{equation}\label{106}
\begin{aligned}
\Psi=\left(e^{3 \Lambda e^{4 \alpha+2 \sqrt{3} \beta_{-}-2 \beta_{+}}}+e^{3 \Lambda e^{4 \alpha-2 \left(\sqrt{3}
   \beta_{-}+\beta_{+}\right)}}\right) e^{-\frac{1}{3} e^{2 (\alpha+\beta_{+})} \cosh \left(2 \sqrt{3}
   \beta_{-}\right)}.
\end{aligned}
\end{equation}

In order to see what effects an aligned electromagnetic field has on our Bianchi VII$_{h=0}$ wave function we will group average the semi classical wave function constructed using (\ref{103}) despite the fact that the addition of an aligned electromagnetic field breaks the reflection invariance in $\beta_{-}$ of the potential. Doing so gives us (\ref{107}). We are doing this so we can make a clear comparison between what happens when we just have a cosmological constant vs when we have both a cosmological constant and an aligned electromagnetic field. Specifically we want to highlight how the aligned electromagnetic field effects the anisotropy present in the universes represented by our wave functions. The actual wave function which has an aligned electromagnetic field would only have one branch present. 

\begin{equation}\label{107}
\begin{aligned}
\Psi=\left(e^{3 \Lambda e^{4 \alpha-2 \left(\sqrt{3} \beta_{-}+\beta_{+}\right)}+\sqrt{3} \beta_{-} \text{b}^2}+e^{3 \Lambda e^{4 \alpha+2 \sqrt{3} \beta_{-}-2 \beta_{+}}}\right) e^{ \left(\frac{1}{6} \left(-2 e^{2 (\alpha+\beta_{+})} \cosh \left(2 \sqrt{3}
   \beta_{-}\right)-3 \text{b}^2 \left(\sqrt{3} \beta_{-}+3 \beta_{+}\right)\right)\right)}
\end{aligned}
\end{equation}

\begin{figure}[!ht]
\centering
\begin{subfigure}{.4\textwidth}
  \centering
  \includegraphics[scale=.165]{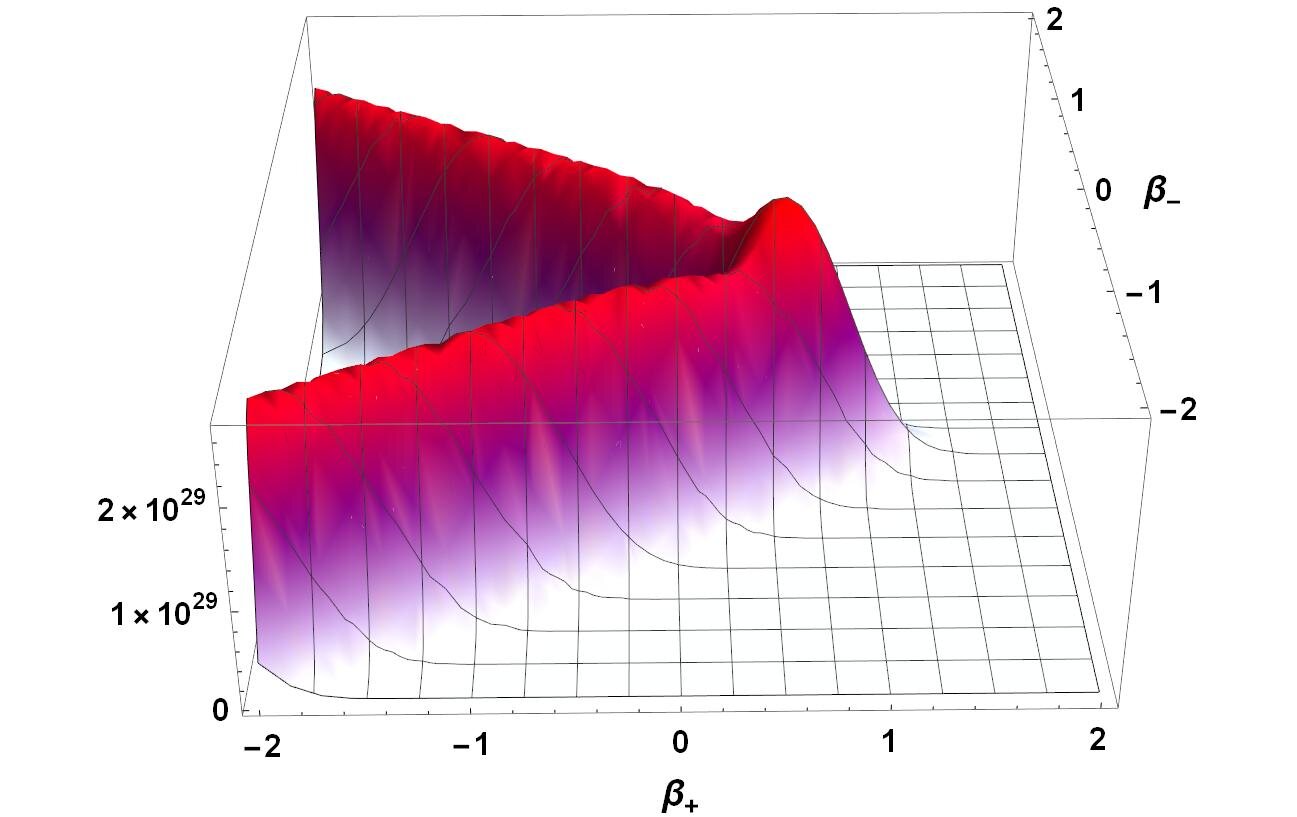}
  \caption{$\alpha=0$ $\hspace{1mm}$ $\Lambda=-1$ $\hspace{1mm}$ $b=0$} 
  \label{7a}
\end{subfigure}%
\begin{subfigure}{.4\textwidth}
  \centering
  \includegraphics[scale=.165]{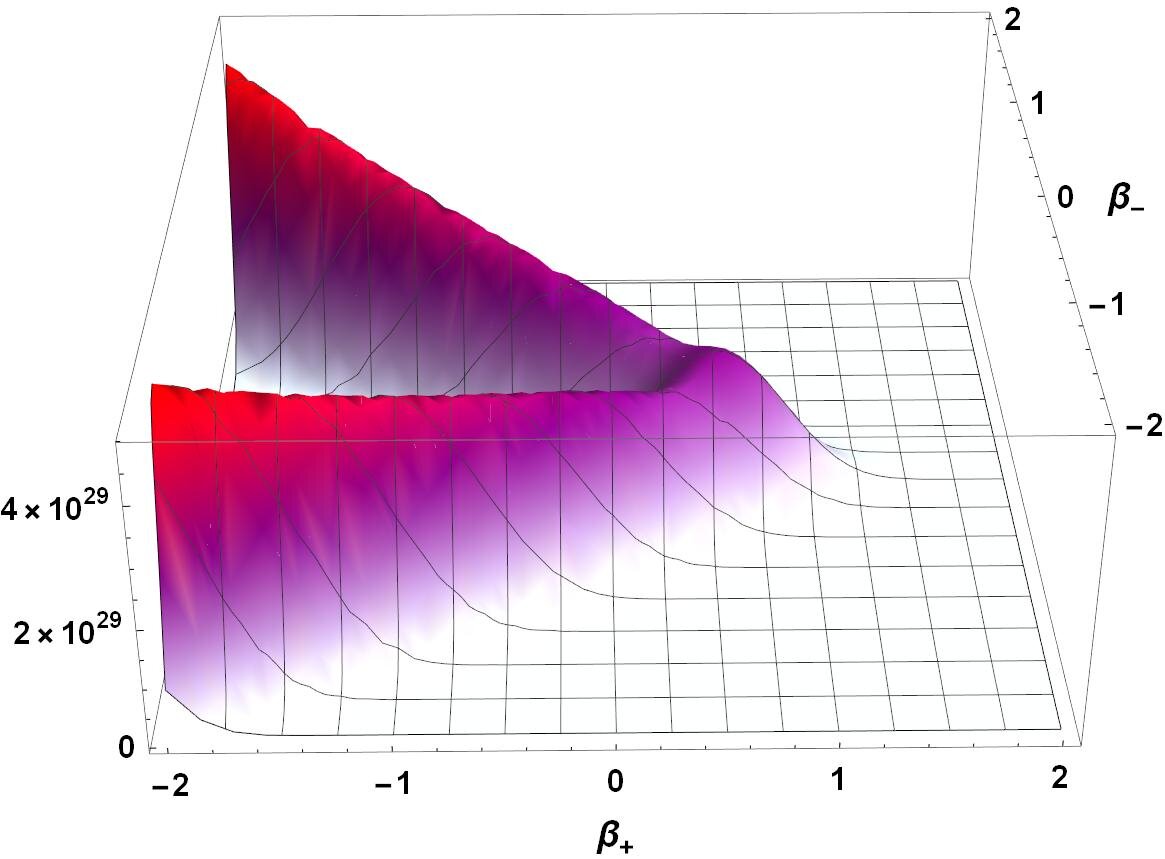}
 \caption{$\alpha=0$ $\hspace{1mm}$ $\Lambda=-1$ $\hspace{1mm}$ $b=\frac{2}{5}$} 
  \label{7b}
\end{subfigure}
\caption{Plots of (\ref{107}) for two different values of the electromagnetic field $b$. Due to the aligned electromagnetic field breaking the reflection symmetry under $\beta_{-}$ of the Bianchi VII$_{h=0}$ potential the actual wave function only has one ridge present as opposed to two, as is shown in figure 20(b). Both ridges are plotted so we can better compare the two wave functions.}
\label{fig 7}
\end{figure}

\section{\label{sec:level1}Discussion }

To begin analyzing our results we first need to adopt an interpretation for the wave functions we computed. Two interpretations of quantum mechanics which in the past have been used to extrapolate physics from Wheeler Dewitt wave functions within the context of quantum cosmology are the consistent histories approach \cite{griffiths1984consistent} and pilot wave theory \cite{bohm1952suggested}. However, for our purposes, we will use the following admittingly naive interpretation which we will briefly outline. Even though we cannot interpret $|\psi|^{2}$ as a probability density due to the lack of a known dynamical unitary operator for the symmetry reduced Wheeler Dewitt equation, if we fix $\alpha$, and only consider wave functions which do not approach $\infty$ when their Misner variables approach $\pm \infty$, our wave functions are reminiscent of normalizable probability densities as can be seen from our plots. Each point in those plots at a fixed $\alpha$ represents a potential geometric configuration the universe can possess which is specified by the values of the Misner variables $\left(\alpha,\beta_+,\beta_-\right)$. Associated with each of those points in $\beta$ space at a fixed $\alpha$ is a value of $|\psi|^{2}$; it is not unreasonable to conjecture that the greater the value of $|\psi|^{2}$ is, the more likely a universe will possess the geometry given by the $\beta_+$ and $\beta_-$ Misner variables. For example if $|\psi\left(\alpha,\beta1_+,\beta1_-\right)|^{2} > |\psi\left(\alpha,\beta2_+,\beta2_-\right)|^{2}$ we would interpret this to mean that a Bianchi II or VII$_{h=0}$ universe described by $\psi$ when it reaches a size dictated by $\alpha$ is more likely to have a spatial geometry which possesses a level of anisotropy described by the values of $\left(\beta1_+,\beta1_-\right)$ as opposed to $\left(\beta2_+,\beta2_-\right)$.

A shortcoming of our interpretation is that it cannot assign numerical values of probability to a micro ensemble of universes with different values of $\alpha$, because $\int^{\infty}_{\infty}\abs{\psi}^{2} d\beta_+ d\beta_-$ is not conserved in $\alpha$. Nonetheless we are picking this interpretation because it is intuitive for the solutions we are dealing with and facilitates the elucidation of the points the author wishes to make. In essence the author was inspired to pick this interpretation because he would like to let the bare solutions speak for themselves. The author strongly encourages future work to be done in extrapolating physics for the results presented in this paper using both the Bohemian approach and consistent histories, in conjunction to other quantitative approaches. What follows will be an attempt to extrapolate the physical implications of our matter sources on the development of quantum Bianchi II and VII$_{h=0}$ universes from their wave wave functions by examining their aesthetic\cite{moncrief1991amplitude} characteristics. We interpret the peaks that appears in our wave functions the same way they are interpreted in \cite{garcia2002noncommutative}, as possible geometric states a quantum universe can tunnel in and out of.

The plots for our 'ground' state asymptotic wave functions in figures (\ref{1a}-\ref{1c}) depict a wave travelling down the $\beta_+$ axis in the negative direction while the scale factor $e^{\alpha}$ grows. As our clock $\alpha$ "ticks" forward in "time", this universe's geometric probability density will be peaked at an ever increasing negative value of $\beta_{+}$ and a roughly constant value of $\beta_{-}$. A variety of other asymptotic wave functions which behave differently could also be constructed using the methods outlined in III. 

Moving on to our Bianchi II solutions with matter sources we notice in figures (\ref{2a}-\ref{2b}) that the aligned electromagnetic field $\left(b^{2}\right)$ causes our wave function to travel in the negative $\beta_{-}$ direction. It was noted in \cite{madsen1989magnetic} that magnetic fields within the context of LRS Bianchi I quantum cosmology induce anisotropy. This is clearly being mirrored for our particular case of the quantum Bianchi II models. However, the opposite can happen as well if our wave functions are centered on the positive portion of the $\beta_{-}$ axis because increasing the strength of the aligned electromagnetic field would initially decrease anisotropy by making our $\beta_{-}$ approach the origin for certain values of $\alpha$ which we take to be our internal clock. There is mathematical evidence that this dual nature of the aligned electromagnetic field is a generic feature possessed by wave functions computed using this method as can be seen in the Bianchi IX models\cite{berkowitz2021bianchi} and Taub\cite{berkowitz2020towards} models. If an aligned electromagnetic field was capable of doing something similar in our early universe then this dual behavior might have been responsible for producing some recognizable signature in the CMB. More research into what signatures a primordial magnetic field could have imparted on the CMB within the context of quantum cosmology is needed. Nonetheless these findings are a good start. 

In addition to inducing anisotropy the electromagnetic field also makes our wave functions thinner and thus more sharply peaked. This effect of making the wave function of the universe more sharply peaked could have played an important role in the early universe by causing quantum states which otherwise would be geometrically fuzzy, such as those whose wave function possesses multiple peaks\cite{berkowitz2020towards,berkowitz2021bianchi} to condense to a far more sharply defined state with one narrow central peak. In other words a primordial electromagnetic field might have played an important role in the early universe by facilitating a phase transition from a quantum universe to one that could be adequately described using classical mechanics. 

One last feature to point out about the aligned electromagnetic field is that its effects rapidly diminish as $\alpha$ grows. As can be seen in figures \ref{2c} and \ref{2d} increasing the strength of the aligned electromagnetic field has milder effects on the wave function then it did in figures \ref{2a} and \ref{2b}. This provides a quantum explanation for why an electromagnetic field might have played a large role in the early universe, but played a diminishing role as it grew in size. 

For figure \ref{3c}, in order for it to be of sufficient quality the author had to multiply the wave function by $10^{20}$, in actuality its magnitude is far less than the wave functions in figures \ref{3a} and \ref{3b}. This is to be expected because a negative cosmological constant naively should act as a powerful force of attraction which resists the tendency of a universe to grow in size. We say naively because as previously mentioned, recent work\cite{hartle2012accelerated,mithani2013inflation, hartle2014quantum} has been done showing that phenomena expected of classical cosmologies with a positive cosmological constant can be derived from certain quantum cosmological models which possess a negative cosmological constant. 

Our wave function (\ref{82}) ($\Lambda < 0$) decaying as $\alpha$ grows indicates that the likelihood that this universe will reach a state when $\alpha >>0$ is low. Another feature of our figures (\ref{3a}-\ref{3c}) is that as $\alpha$ grows they become thinner. This shows that as $\alpha$ grows larger the universe that these wave functions are supposed to represent become somewhat less fuzzy. Geometric fuzziness is a feature associated with the uncertainty relation of our minisuperspace as a result of quantum mechanics. As a result the larger a universe becomes the less we expect those quantum features of fuzziness to be present. 

Our leading order 'excited' states when $\Lambda \ne 0$, figures (\ref{4a}-\ref{4d}), can be interpreted similarly to our other states. One noticeable difference is that their geometry is more "fuzzy" because the wave functions which describe this universe have multiple ridges/peaks. This is most clearly illustrated in the $\alpha=0$ case where we can see three distinct ridges, one large ridge and two smaller ridges. Figure \ref{4d} further shows this. The potential geometries this Bianchi II universe can take on are located on one of these three visible ridges. As $\alpha$ grows those ridges appear to fuse and the geometries become slightly more "sharp", while also becoming more unlikely to occur due to the magnitude of the wave function decaying. It is also possible that each one of those ridges represents a "fuzzy" geometric state that our quantum Bianchi II universe can tunnel in and out of. 

For the non-commutative quantum Bianchi II models our wave function behaves similarly to other non-commutative quantum cosmological models \cite{garcia2002noncommutative,socorro2009scalar} in the sense that multiple peaks are present when $\theta_{i} \ne 0$, and only one defined peak is present when $\theta_{i} = 0$. These multiple peaks indicate that non-commutativity in the early universe could facilitate the creation of many possible states which a quantum universe can tunnel into. Our universe might have been one of those states; and if it wasn't for non-commutativity it might have been exceedingly unlikely for our universe to tunnel into the state which allowed it to evolve in the way it has. 

When we include our aligned electromagnetic field in figure \ref{5b} it doesn't have  much of an effect. However because we simplified our wave function by setting all three non-commutative parameters $\theta_{i}$ equal to each other we only looked at a small subset of possible quantum non-commutative universes. There very well could exist a region in $\theta$ space such that a primordial electromagnetic field has a profound effect on what states the early universe can tunnel into. Studying the full effects of both non-commutativity and primordial electromagnetic fields in the early universe could shed much light on how our homogeneous and isotropic universe came to be.  

Our vacuum Bianchi VII$_{h=0}$ 'excited' states(figures \ref{6a}-\ref{6b}) are two Gaussian like peaks which travel in the negative $\beta_{+}$ direction. Thus for a specific value of $\alpha$ our wave function represents a universe which can tunnel in between two relatively defined anisotropic states. This behavior is unusual if compared to that of other\cite{berkowitz2020towards,bae2015mixmaster} 'excited' states computed using this method. Usually for $\alpha < 0$ an 'excited' state possesses multiple peaks that each represent a geometric configuration a universe can tunnel in and out of. However when $\alpha$ grows large, $\alpha >>0$, those peaks merge into a single peak. This process can be interpreted as one in which a spatially small universe which is quantum in nature evolves into a spatially larger universe which behaves more in line with classical physics. This merger of the peaks never happens for our Bianchi VII$_{h=0}$ 'excited' state, the two peaks remain separate for all values of $\alpha$. 

This difference in how the 'excited' states of these models behave may be accounted for by the different properties of the groups that the Bianchi VII$_{h=0}$ and IX models are based on. The Lie algebra corresponding to the Bianchi VII$_{h=0}$ models is the Lie algebra of the group of isometries of the plane, while the Lie algebra of the Bianchi IX models is associated with the rotational SO(3) group associated with the sphere. The mathematical differences between these models underlining groups or their topological differences may account for the different behavior of their 'excited' states for large $\alpha$. By further comparing and contrasting these models and we can learn more about how 'excited' states are defined within the context of this method and in quantum cosmology.

For our quantum Bianchi VII$_{h=0}$ models with matter sources(figures \ref{7a}-\ref{7b}) it can be seen that when the aligned electromagnetic field $b^{2}$ is zero our wave function is peaked around $\beta_{+}=\frac{1}{2}$ and $\beta_{-}=0$. However when our aligned electromagnetic field increases in strength the ridges overtake the aforementioned peak and our wave function represents a fuzzier and more anisotropic universe. Making a wave function of the universe less geometrically defined is another potential effect that an electromagnetic field can have. Within the context of non-commutative quantum cosmology this could mean creating additional peaks which otherwise wouldn't be present. 

\section{\label{sec:level1}Concluding Remarks} 

From a mathematical point of view we showed how useful the Euclidean-signature semi classical method is for tackling problems which possess the same symmetries as those that are present in the vacuum Bianchi II models. In addition we have shown how this method can shed new light on problems which aren't as symmetric as the vacuum Bianchi II models, such as the Bianchi II and VII$_{h=0}$ models when matter sources are included. To the author's knowledge the solutions we presented in this paper for the Bianchi II and VII$_{h=0}$ models are new and are a result of the utility of this method. 

Even for the cases where we were only able to solve for the Euclidean-signature Hamilton Jacobi equation we have shown the usefulness of this approach. As previously mentioned, any valid operator ordering of the Lorentzian signature functional WDW equation can be connected to a functional Einstein-Hamilton-Jacobi equation. The Euclidean-signature semi classical method allows us to relate the Lorentzian functional WDW equation to the functional Euclidean-signature EHJ equation 

\begin{equation}\label{109}
\left(\frac{16 \pi G}{c^{3}}\right)^{2} \frac{\left(\gamma_{i k} \gamma_{j \ell}-\frac{1}{2} \gamma_{i j} \gamma_{k \ell}\right)}{\mu_{\gamma}} \frac{\delta S_{(0)}}{\delta \gamma_{i j}} \frac{\delta S_{(0)}}{\delta \gamma_{k \ell}}+\mu_{\gamma}^{(3)} R(\gamma)=0,
\end{equation}
where 
$\gamma_{i k}$ is the Riemannian metric induced on the spatial 3-hypersurface by $g_{\mu \nu}$, $\mu_{\gamma}$ is the determinant of $\gamma_{i k}$, and $^{(3)} R(\gamma)$ is the Ricci scalar of $\gamma_{i k}$. As a result of this being a Euclidean-signature equation we may be able to prove the existence of solutions to it by appealing to theorems that apply to Euclidean-signature geometries such as the positive action theorem \cite{schoen1979complete,schoen1979proof,zhang1999positive,dahl1997positive}. More information on this program in so far as it applies to quantum gravity can be found in section 7 of \cite{moncrief2014euclidean}. By finding new solutions to the Euclidean-signature Hamilton Jacobi equations of the Bianchi II and VII$_{h=0}$ models we have provided further support for the idea that Euclidean-signature equations are easier to handle than their Lorentzian signature counterparts, thus providing further incentive to advance the development of this method.

An attractive feature of this method is that it doesn't rely on using a Wick rotation to connect the solutions of its Euclidean-signature equations to solutions for the Lorentzian signature equations it is applied to. This makes it an attractive method for tackling certain problems in bosonic field theory and quantum gravity. For bosonic field theory in particular the Euclidean-signature semi classical doesn't require splitting the theory up into one part which is linear(non-interacting) and another part which is a nonlinear(interacting) perturbation. This allows the fully interacting nature of the field theory to be present\cite{marini2019euclidean,marini2020euclidean} at every level of its analysis. Additionally we extended the applicability of this technique by successfully employing it when all of the conditions\cite{bae2015mixmaster} that we previously outlined were not present.

In terms of pure quantum cosmology, in our paper we found many new solutions to the Wheeler DeWitt equations for the quantum Bianchi II and VII$_{h=0}$ models when a cosmological constant, an aligned electromagnetic field and a stiff matter term were present. By doing so we greatly expanded upon the known number of closed form solutions to the Wheeler Dewitt equation and provided results which can shed light on how matter sources affect the evolution of a quantum universe. Notably they further point\cite{berkowitz2021bianchi,berkowitz2020towards} to a whole host of generic behaviors that are imparted to the wave function of the universe from a primordial aligned electromagnetic field. In order for the results in this paper to be potentially more relevant to our own universe in its early infancy it is crucial that the analysis that we carried out be done for general electromagnetic field configurations found in non-diagonal Bianchi models and include inhomogeneous perturbations. This is something that will be pursued in the future. We can also expand our results in the future by including a scalar field.

\section{\label{sec:level1}ACKNOWLEDGMENTS}
 
I am grateful to Professor Vincent Moncrief for valuable discussions at every stage of this work. I would also like to thank George Fleming for facilitating my ongoing research in quantum cosmology. Daniel Berkowitz acknowledges support from the United States Department of Energy through grant number DE-SC0019061. I also must thank my aforementioned parents.

\bibliography{Bianchi}

\end{document}